\documentclass[natbib,epsfig,graphicx]{jfm}

\usepackage{amsmath}
\usepackage{bm}
\usepackage{graphicx}
\usepackage{natbib}
\usepackage{tikz}
\usepackage{amsmath}
\usepackage{subfigure}
\usepackage{comment}
\usepackage{enumerate}
\usetikzlibrary{%
    arrows
}

\ifCUPmtlplainloaded \else
  \checkfont{eurm10}
  \iffontfound
    \IfFileExists{upmath.sty}
      {\typeout{^^JFound AMS Euler Roman fonts on the system,
                   using the 'upmath' package.^^J}%
       \usepackage{upmath}}
      {\typeout{^^JFound AMS Euler Roman fonts on the system, but you
                   dont seem to have the}%
       \typeout{'upmath' package installed. JFM.cls can take advantage
                 of these fonts,^^Jif you use 'upmath' package.^^J}%
      }
  \else
  \fi
\fi

\ifCUPmtlplainloaded \else
  \checkfont{msam10}
  \iffontfound
    \IfFileExists{amssymb.sty}
      {\typeout{^^JFound AMS Symbol fonts on the system, using the
                'amssymb' package.^^J}%
       \usepackage{amssymb}%

      }{}
  \fi
\fi

\ifCUPmtlplainloaded \else
  \IfFileExists{amsbsy.sty}
    {\typeout{^^JFound the 'amsbsy' package on the system, using it.^^J}%
     \usepackage{amsbsy}}
    {\providecommand\boldsymbol[1]{\mbox{\boldmath $##1$}}}
\fi

\newcommand{\suchthat}{\;\ifnum\currentgrouptype=16 \middle\fi|\;}

\newsavebox{\astrutbox}
\sbox{\astrutbox}{\rule[-5pt]{0pt}{20pt}}

\title[Zero-pressure-gradient turbulent boundary layer]{Symmetries and similarities of the zero-pressure-gradient turbulent boundary layer}
\author[C.~Tong]%
{C\ls H\ls E\ls N\ls N\ls I\ls N\ls G\ns T\ls O\ls N\ls G
\thanks{Email address for correspondence: ctong@clemson.edu}
}

\affiliation{Department of Mechanical Engineering, Clemson University, Clemson, SC 29634, USA}%\\[\affilskip]}

\begin{document}

\maketitle

\begin{abstract}
  The symmetries and similarities of the zero-pressure-gradient turbulent  boundary layer (ZPGTBL) are investigated
  to derive the full set of similarity variables, to derive the similarity equations, and to obtain a higher-order approximate solution
  of the mean velocity profile. \textcolor{black}{Previous analyses have not resulted in all the similarity variables.}
  % while experimental works have  been based largely on the wake function, which lacks an analytic origin.}
  We perform a symmetry analysis of the equations for ZPGTBL using Lie dilation groups, and obtain local, leading-order symmetries of
  the equations. The full set of similarity variables were obtained in terms of the boundary layer parameters. The friction velocity was shown to be the
  outer-layer velocity scale.
  The downstream evolutions of the boundary thickness and the friction velocity are obtained analytically.
The dependent similarity variables
are written as asymptotic expansions.% to derive the perturbation equations for both the outer and inner layers.
%The gauge functions are shown to be different from those in channel flows at the higher orders.
By asymptotically matching the expansions, an approximate similarity solution up to the third order in the overlapping layer are obtained.
  These results are obtained from first principles without any major assumptions \textcolor{black}{and a turbulence model.}
The similarities and differences between ZPGTBL and turbulent channel flows in terms of the similarity equations, the gauge functions
and the approximate solutions are discussed. In particular, the leading-order expansions are identical for ZPGTBL and channel flows, supporting 
the notion of universality of the near-wall layer. \textcolor{black}{In addition, the logarithmic frictionlaw for ZPGTBL is accurate to all orders while it is only accurate
at the leading order in channel flows.} The results will help further understand ZPGTBL and
the issue of universality of the near-wall layer in wall-bounded turbulent flows. 
\end{abstract}

%\begin{keywords}
%Authors should not enter keywords on the manuscript, as these must be chosen by the author during the online 
%submission process and will then be added during the typesetting process 
%(see http://journals.cambridge.org/data/\linebreak[3]relatedlink/jfm-\linebreak[3]keywords.pdf for the full list)
%\end{keywords}

%\newpage
\baselineskip 20pt
\parskip 6pt
\section{Introduction}

%Turbulent boundary layers are of great fundamental and practical interest. The most basic of such flows is
The zero-pressure-gradient turbulent  boundary layer (ZPGTBL) is the most fundamental of turbulent boundary layers, and has been studied extensively
(e.g., \citealt{Klebanoff1954,Schlichting1956,Clauser1956,MY71,Sreenivasan1989,Pope2000,McKeonSreenivasan2007,Nagib2007,Marusic2010,Smits2011}).
The flow consists of two layers with different scaling. The inner layer has viscous scaling and the mean velocity 
follows the law of the wall, and is considered universal (\citealt{Prandtl1925}). The outer layer does not explicitly depend on the viscosity and
the mean velocity obeys the
velocity defect law (\citealt{vK1930}). There is a matching sublayer between the outer and inner layers where the mean velocity is generally considered to
follow the log law (\citealt{vK1930}).

The most important fundamental understanding of a boundary layer is its similarity equations and similarity solution.
The zero-pressure-gradient laminar boundary layer (the Blasius boundary layer, \citealt{Blasius1908}) has the well-known similarity solution.
By expressing the variables in the partial differential equations
for the boundary layer in terms of the similarity variables, the equations were reduced to an ordinary differential equation, the solution of which is
the similarity solution. The downstream evolutions of the boundary layer thickness and the surface stress were obtained analytically.
By contrast, there has not been a similarly successful development for ZPGTBL to date, despite of the extensive past research devoted to the topic.
\cite{TL72} were the first to obtain the leading-order mean momentum similarity equation and to use it to predict the log law. However,
the boundary layer thickness $\Delta$ was defined by \cite{Clauser1956} using an integral quantity, rather than the boundary layer parameters. In addition,
its downstream evolution as well as that of the friction velocity was not predicted. 
With this definition of $\Delta$,  the non-dimensional outer layer wall-normal coordinate, the independent similarity variable,
was not fully defined, in contrast to the Blasius boundary layer, making it difficult to derive the higher-order similarity equations and
to analyse the similarities. \cite{OK2006} performed a Lie group analysis of the two-point correlation and found a linear growth of $\Delta$.
\cite{Monkewitz2007} used the same definition of the boundary layer thickness and
derived an expression for the downstream evolution of the momentum thickness, which depends on the shape factor $H$, while
no expression for $\Delta$ was provided. They also fitted experimental data to asymptotic expansions with the gauge functions depending on
the friction Reynolds number.
\cite{George1997} used the Reynolds-averaged boundary layer equations without the viscous terms to obtain a similarity solution.
{ However, as will be shown in the present work, by eliminating the viscous stress terms alone, the equations do not have a similarity
solution as a function of the local wall-normal coordinate.}

In the meantime, much progress in understanding ZPGTBL has come from experiments. 
\cite{Coles1956} proposed the wake function empirically based on experimental data, which is the departure from the log law in the outer layer.
The combination of the two is proposed as the outer layer similarity profile.
Much of the later work on the outer layer similarities
has been based on the wake function, e.g., \cite{Perry1994}, \cite{Nagib2007}, \cite{McKeon2007}, \cite{Marusic2015} and \cite{Vallikivi2015a}.
Nevertheless, the wake function lacks an analytic origin.
%For example, Perry 1994 used the combined log law and wake function as the similaity solution to
%obtain expressions for the Reynolds shear stress.
%and leads to results inconsistent with the known properties of ZPGTBL. 
%to date there has not been a similar successful reduction of the
%Reynolds-averaged boundary layer equations, which include the Reynolds-averaged Navier Stokes equations and the Reynolds stress budgets,
%to obtain ordinary differential equations describing a similarity solution.
%The similarity variables have not been fully defined, e.g., the non-dimensional outer layer
%vertical coordinate, because the non-dimensional boundary layer thickness is not defined in term of the flow parameters, $x$, $u_*$, $U_e$. 
To gain a comprehensive understanding of the similarities of ZPGTBL, in the present work we will perform a symmetry analysis of the
Reynolds-averaged boundary layer equations to derive analytically the full set of similarity variables,
the evolution of the boundary layer thickness, the evolution of the friction velocity and
the ordinary differential equations describing the similarity solution.
We will then employ the method of matched asymptotic expansions obtain a higher-order approximate solution. To our best knowledge, this is
the first time such an analysis has been carried out.

Parallel to the research on ZPGTBL, there also have been much effort to investigate turbulent channel and pipe flows, which are 
amenable to more rigorous asymptotic analysis (e.g., \citealt{Millikan38,Afzal1976}), including higher-order corrections to the log law.
The near-wall, ``constant stress'' layer in ZPGTBL is commonly believed to have much in common with channel flows,
i.e., the near-wall layers  in these wall flows are believed to be universal, with the log law as the leading-order velocity profile.
%The argument leading to the log layer (\citealt{TL72}) is largely identical to that for channel flows.
There is also evidence against universality (e.g., \citealt{McKeon2007,Nagib2007,Marusic2015}).
However, there has been essentially no theoretical analysis on the similarities and differences regarding the Reynolds number dependence between these
flows, e.g., no higher-order asymptotic expansions for ZPGTBL to be compared to those for the channel flows.
The present work will help shed some light on the important issue of the universality of the near-wall layers.

There have also been arguments questioning the inner-layer similarity among these flows. \cite{George1997} claimed that the mean
advection terms in the momentum equation for ZPGTBL fundamentally changes the scaling of the outer layer
mean velocity profile. They argued that the velocity defect scales with the free-stream velocity $U_e$ rather than  the friction velocity $u_*$.
This scaling also leads to a power-law velocity profile in the overlapping layer, which is not universal, with Reynolds number dependent
exponent and coefficients. On the other hand, \cite{Jones2008} argued that both $U_e$ and $u_*$ are valid scaling candidates.
In the present work we show that same as in channel and pipe flows,
the friction velocity is the correct scale for the velocity defect in ZPGTBL. 

The rest of the paper is organized as follows. In Section 2 we perform a symmetry analysis using Lie dilation groups and obtain
the full set of similarity variables. We then derive the similarity equations as singular perturbation equations, obtain an approximate solution
as asymptotic expansions and make some comparisons with measurements in Section 3, followed by Discussions and Conclusions.

\section{Symmetry analysis}

The Naiver-Stokes equations have a number of symmetries. One of them
is that the equations are invariant under a one-parameter Lie dilation group (e.g., \citealt{Bluman1989}, \citealt{Frisch1995}, \citealt{Cantwell2002}),
which is closely related to
the scaling properties and the similarity of the solution. Since invariance of the equations under the group transformation requires
a constant Reynolds number, full similarity of the solution is
possible only when the Reynolds number is fixed. However, it has long been recognized that the energy-containing and flux-carrying
statistics in turbulent flows at high Reynolds numbers are approximately Reynolds number invariant. The leading-order flow behaviors,
including similarity properties, are approximately independent of the Reynolds number.
This approximate Reynolds number invariance is associated with spontaneous breaking
of symmetries of the Navier-Stokes equations from laminar to turbulent flows as the Reynolds number increases.
Therefore, while the symmetries of laminar flows are exact, the symmetries
of turbulent flows are only approximate. The concept of spontaneous symmetry breaking and approximate symmetry first emerged in condensed matter
physics and later were key to predicting certain non-zero mass particles in Yang-Mills gauge fields
(\citealt{Higgs1964,Higgs2014,Castellani2003}). %, since at high temperatures the fields have exact symmetries with massless particles (\citealt{Goldstone1962}).
%Self-similarity of some evolving flows (e.g., plane turbulent jets) is a result of the Reynolds number invariance, although they are not
%strickly self-self similar because the Reynolds number changes as they evolve in space or time.
 As we will show in the following, the symmetries of ZPGTBL
 also can only be at the leading order and be approximate.
 %and Reynolds number independent property, with any Reynolds number dependence being of higher orders.
 In the present work, we seek the leading-order symmetries and similarity properties and determine the higher-order corrections
to account for any Reynolds-number dependence.

We use Lie dilation groups to analyse the symmetries of the Reynolds-averaged boundary layer equations. % to identify their similarity properties.
Similar to the Navier-Stokes equations, the boundary layer equations
also have certain group transformation properties. Once found, the group can be
used to determine the similarity variables of the equations. Then using the similarity variables, the partial differential
boundary layer equations can be reduced to ordinary differential equations, whose solution
is the similarity solution. For ZPGTBL, the equations are the mean momentum equation and the
Reynolds stress budget (or the shear-stress budget and the turbulent kinetic budget for our analysis). We need to identify the Lie dilation group under which
all three in this set of equations are invariant.

We now analyse the Lie dilation group of the equations. %, which is closely related to the scaling properties of the equations.
The Reynolds-averaged momentum equation for ZPGTBL can be written in a form similar to that given by \cite{TL72}
\begin{equation} \label{full_momentum}
 U\frac{\partial U}{\partial x}+V\frac{\partial U}{\partial y}=-\frac{\partial{\overline{uv}}}{\partial{y}}
  -\frac{\partial}{\partial x}(\overline{u^2}-\overline{v^2}) + \nu\frac{\partial^2 U}{\partial y^2},
\end{equation}
where $U$, $V$, $\overline{uv}$, $\overline{u^2}$, $\overline{v^2}$, and $\nu$ are the streamwise and normal mean velocity components, the Reynolds shear stress,
the Reynolds normal stress components and the kinematic viscosity, respectively.
The finite form of the dilation group is
\begin{align} \label{eq_dilation1}
\begin{split}
 & \tilde{x}=e^ax,\ \tilde{y}=e^by, \ \tilde{U}=U,  \ \tilde{U}-U_e=e^g(U-U_e),  \ \partial\tilde{ U}=e^g\partial U, \ \tilde{V}=e^cV,\\
 &  \widetilde{\overline{uv}}=e^d\overline{uv}, \ \widetilde{\overline{u^2}}=e^{d_1}\overline{u^2},\ \widetilde{\overline{v^2}}=e^{d_1}\overline{v^2}.
\end{split}
\end{align}
where $a$, $b$ etc., are the group parameters.
Note that the mean velocity can dilate differently than statistics of the velocity fluctuations, which has been
pointed out by \cite{She2017} based on the argument of random dilation groups.
\textcolor{black}{In the context of ZPGTBL, dilation can be thought of as rescaling of the physical variables for the given free-stream velocity $U_e$ and $\nu$.}
The streamwise velocity $U$ in this problem \textcolor{black}{cannot} dilate because the free-stream velocity, once specified, is fixed.
However, the velocity defect and the velocity differential $\partial U$ can dilate \textcolor{black}{(e.g, dilate along with $u_*$ and $y$).
  We will examine the possibilities that $U-U_e$ and $\partial U$ do and do not dilate.
  Here we first examine the symmetry if they not dilate.}  Substitute \ref{eq_dilation1} into \ref{full_momentum}
\textcolor{black}{and drop the tildes in the equations for convenience hereafter}, we have
\begin{equation}
  e^{-a}U\frac{\partial U}{\partial x}+e^{c-b}V\frac{\partial U}{\partial y}=-e^{d-b}\frac{\partial{\overline{uv}}}{\partial{y}}
  -e^{d_1-a}\frac{\partial}{\partial x}(\overline{u^2}-\overline{v^2}) + e^{-2b}\nu\frac{\partial^2 U}{\partial y^2}.
\end{equation}
For the equation to be invariant under the dilation group, the exponents for each term must be equal:
\begin{equation}
  -a=c-b=d-b=d_1-a=-2b.
\end{equation}
Therefore
\begin{equation} \label{eq_exponents2}
  a=2b, \ \ c=-b, \ \ d=-b,\ \  d_1=0.
\end{equation}
Thus \ref{full_momentum} with the boundary condition ($U=U_e$ for $y \rightarrow \infty$) is invariant under a one-parameter dilation group,
\begin{equation}
  \tilde{x}=e^{2b}x, \ \tilde{y}=e^by, \ \tilde{U}=U, \ \partial\tilde{ U}=\partial U, \ \tilde{V}=e^{-b}V, \ \widetilde{\overline{uv}}=e^{-b}\overline{uv}.
\ \widetilde{\overline{u^2}}=\overline{u^2},\ \widetilde{\overline{v^2}}=\overline{v^2},
\end{equation}
A one-parameter dilation group fully determines the scaling of the variables in the problem and can be used to identify the similarity variables.
The group specified by \ref{eq_exponents2} is essentially the same as that for the zero-pressure-gradient laminar (Blasius) boundary layer
(\citealt{Cantwell2002}). It ($a=2b$)
indicates a growth rate for the boundary layer thickness $\delta \sim x^{\frac{1}{2}}$, the same as the Blasius boundary layer, but is
inconsistent with that of the turbulent boundary layer.
We further consider the shear-stress budget,
\begin{equation} \label{uv_budget}
  U \frac{\partial  \overline{uv}}{\partial x}+  V\frac{\partial  \overline{uv}}{\partial y}=
  -\overline{u\frac{\partial p}{\partial y}+v\frac{\partial p}{\partial x}}-  \overline{v^2}\frac{\partial U}{\partial y}
  -\frac{\partial  \overline{uv^2}}{\partial y}+\nu \frac{\partial ^2 \overline{uv}}{\partial y^2}-\epsilon_{uv},
\end{equation}
where $\epsilon_{uv}$ is the dissipation rate, which is generally negligible.
For this equation to be invariant, the exponents for the terms on the l.h.s and the second term on the r.h.s. must satisfy
\begin{equation}
  d-a=c+d-b=d_1-b,
\end{equation}
which leads to $d=a-b$, inconsistent with \ref{eq_exponents2}. Therefore, there is no dilation group under which both equations \ref{full_momentum} and
\ref{uv_budget} are invariant. Consequently, there is no full similarity solution for which the velocity defect scales as the $U_e$.

\textcolor{black}{We now examine the symmetry if the velocity defect and the velocity differential dilate as}
\begin{equation} \label{}
  \tilde{U}-U_e=e^g(U-U_e),  \ \partial\tilde{ U}=e^g\partial U,
\end{equation}
\textcolor{black}{The streamwise mean advection term then can be written as
\begin{equation} \label{advection}
  U\frac{\partial U}{\partial x}=U_e\frac{\partial U}{\partial x}+(U-U_e)\frac{\partial U}{\partial x}.
  \end{equation}
The exponents must satisfy
\begin{equation}
  g-a=2g-a,
\end{equation}
resulting in $g=0$, the velocity defect cannot dilate.}
Therefore, as in the case of the (full) Navier-Stokes equations for a general turbulent
flow, there are no exact symmetries for the full ZPGTBL equations. Only approximate symmetries
are possible. In seeking the approximate symmetries, we recognise that viscous effects are negligible in the outer layer whereas they play a leading-order role
in the inner layer, indicating that the approximate symmetries are not global, but local. In the following we analyse the symmetries of the
outer and inner layers separately. 

\subsection{Outer layer symmetry}

For the outer layer, the leading-order symmetries or the symmetries of the leading-order equations can be obtained by dropping the Reynolds-number-dependent
terms. These symmetries similar to the approximate symmetries of the Navier-Stokes equations previously investigated (e.g., \citealt{Grebenev2007}).
We first drop the viscous term in the mean momentum equation, which is explicitly Reynolds-number dependent and is a higher-order term.
\textcolor{black}{We note that as shown above, due to the (full) advection term (\ref{advection}) $U-U_e$ and $\partial U$ do not dilate.}
For the momentum equation to be invariant under the dilation group, the exponents must satisfy,
\begin{equation}
  -a=c-b=d-b=d_1-a.
\end{equation}
Therefore
\begin{equation}
  c=b-a, \ \ d=b-a, \ \  d_1=0.
\end{equation}
The mean momentum equation is therefore invariant under the two-parameter dilation group
\begin{equation}
  \tilde{x}=e^{a}x, \ \tilde{y}=e^bx, \ \tilde{U}=U, \ \partial \tilde{U}=\partial U, \ \tilde{V}=e^{b-a}V, \ \widetilde{\overline{uv}}=e^{b-a}\overline{uv},
\ \widetilde{\overline{u^2}}=\overline{u^2}, \ \widetilde{\overline{v^2}}=\overline{v^2},
\end{equation}
which does not fully determine the scaling of the variables. To determine the scaling we need to identify a one-parameter group by
relating the parameters $a$ and $b$. To do so we further consider the shear-stress budget with the viscous terms dropped.
%\begin{equation}
%  uv.
%\end{equation}
For this equation to be invariant, the exponents must satisfy
\begin{equation}
  b-2a=2b-2a-b=-b.
\end{equation}
Therefore
\begin{equation} \label{eq_ab}
  a=b,
\end{equation}
resulting in a one-parameter dilation group. However, this group indicates that the Reynolds  shear stress does not dilate, i.e.,
it is a constant,
and that  $\delta$ grows linearly, inconsistent with the known behaviours of the turbulent boundary layer.
Therefore, dropping the explicitly Reynolds-number-dependent viscous terms {\it alone}, as done by \cite{George1997}, does not lead to the correct local symmetries.
The reason is that there are other higher-order terms in the equations that do not contain the viscosity, but are implicitly Reynolds-number dependent.
They also need to be identified and dropped in order to obtain the leading-order outer-layer symmetries and the leading-order outer similarity solution.

To identify the higher-order terms, we perform an order of magnitude analysis of the equations. Again we first consider the scaling choice
$U-U_e\sim U_e$.
%The leading-order mean momentum equation is
%\begin{equation}
%U_e \frac{\partial U}{\partial x}+V\frac{\partial U}{\partial y}=-\frac{\partial \overline{uv}}{\partial y},
%\end{equation}
%where the terms scale as $u_*^2/\delta$. The scaling of $V$ is obtained from the mean continuity equation. 
%\begin{equation}
%  U_e \frac{\partial  \overline{uv}}{\partial x}=-\overline{u\frac{\partial p}{\partial y}+v\frac{\partial p}{\partial x}}-
%  \overline{v^2}\frac{\partial U}{\partial y}.
%\end{equation}
With this choice, the shear production of the TKE scales as $-\overline{uv}(\partial U/\partial y) \sim u_*^2U_e/\delta$. Since it is the only production term,
the TKE scales as $k\sim U_e^2$. The velocity variances scale the same way.
The dissipation rate would scale as $k^{3/2}/\delta\sim U_e^3/\delta$ (\citealt{Taylor1935}), asymptotically larger than the production rate,
indicating that the scaling choice $U-U_e\sim U_e$ is inconsistent with the scaling of the dissipation rate. Therefore, the velocity
defect cannot scale as $U_e$. % as the scaling is inconsistent with the dynamics of the TKE.

With $U_e$ ruled out, the only scaling choice for the velocity defect is $U-U_e\sim u_*$, as there are no other velocity scales
in the problem. Again, to obtain
the one-parameter dilation group that is consistent with the leading-order local symmetries 
 of the outer layer, additional terms that do not contain $\nu$, but still
depend on the Reynolds number must be identified and dropped.
We perform an order of magnitude analysis of the mean momentum equation and the shear stress budget to identify the leading-order terms,
which are Reynolds-number independent. The orders of magnitude of the terms in the mean momentum equation are

\begin{align} \label{eq_MM_order}
\begin{split}
& U_e\frac{\partial U}{\partial x}+(U-U_e)\frac{\partial U}{\partial x}\hspace{0.2cm}+\hspace{0.25cm} V\frac{\partial U}{\partial y}\hspace{0.25cm}=\hspace{0.25cm}-\frac{\partial \overline {uv}}{\partial y}-\frac{\partial(\overline{u^{2}}-\overline{v^{2}})}{\partial x}+\nu\frac{\partial^{2}U}{\partial y^{2}}\\
&\ U_e\frac{u_*}{L} \hspace{0.8cm} \frac{u_*{^2}}{L} \hspace{1.5cm}  V\frac{u_*}{\delta}=\frac{u_*^{2}}{L} \hspace{0.8cm} \frac{u_*^{2}}{\delta} \hspace{1.2cm} \frac{u_*^{2}}{L} \hspace{1.30cm} \nu\frac{u_*}{\delta^{2}}\\
&\  \hspace{0cm} O(1) \hspace{0.8cm} O(\frac{u_*}{U_e}) \hspace{1.45cm} O(\frac{u_*}{U_e}) \hspace{1.2cm} O(1) \hspace{0.8cm} O(\frac{u_*}{U_e}) \hspace{0.7cm} O(Re_*^{-1}), 
\end{split}
\end{align}
where the streamwise and wall-normal length scales are $L$ and $\delta$ respectively. 
The advection due to $U_e$ must be of $O(1)$ to balance the Reynolds shear stress gradient, resulting in $U_e/L\sim u_*/\delta$.
The mean continuity equation is used to obtain $V \sim \delta u_*/L \sim u_*^2/U_e$.
Multiplying the equation by $\delta/u_*^{2}$ gives the orders of magnitude in the third line in \ref{eq_MM_order}. 
The orders of magnitude of the terms in the Reynolds shear stress budget are
%\begin{align}
 %   \begin{split}
  %      U_e\frac{u_*}{L} \sim \frac{u_*^{2}}{\delta}, \frac{L}{\delta}=\frac{U_e}{u_*}, L\sim x\\
   %     &\ \hspace{-4.3cm} v \sim \frac{\delta u_*}{L}, V\frac{\partial U}{\partial y}\sim\frac{\delta}{L} U_* \frac{u_*}{\delta}\\
    % &\ \hspace{-4.1cm} \sim \frac{u_*^{2}}{U_e}
 %   \end{split}
%\end{align}
\begin{align}
    \begin{split}
      & U_e\frac{\partial \overline{uv}}{\partial x}+(U-U_e)\frac{\partial \overline{uv}}{\partial x}+V\frac{\partial \overline{uv}}{\partial y}
      =-\overline{(u\frac{\partial p}{\partial y}+v\frac{\partial p}{\partial x})}-\frac{\partial \overline{uv^{2}}}{\partial y}
      -\overline{v^{2}}\frac{\partial U}{\partial y}+\nu\frac{\partial^{2}\overline{uv}}{\partial y^{2}}-\epsilon_{12}\\
      &\ \frac{u_*^{3}}{\delta} \hspace{1.4cm} \frac{u_*^{3}}{L} \hspace{1.75cm} \frac{\delta}{L}\frac{u_*^{3}}{\delta} \hspace{1.5cm}  \frac{u_*^{3}}{\delta} \hspace{2.7cm}
      \frac{u_*^{3}}{\delta} \hspace{1.1cm}
      \nu \frac{u_*^{2}}{\delta^{2}}\\
       &\ O(1) \hspace{1.0cm} O(\frac{u_*}{U_e}) \hspace{1.2cm} O(\frac{u_*}{U_e}) \hspace{1.3cm} O(1) \hspace{2.5cm} O(1) \hspace{0.3cm} \hspace{0.2cm} O(Re_*^{-1}).
    \end{split}
\end{align}
\textcolor{black}{The turbulent transport term (the second on the right-hand side) and the dissipation terms are small but we do not have the exact order of magnitudes.}
Multiplying the equation by $\delta/u_*^3$ gives the orders of magnitude. The orders of magnitude of the TKE budget are
\begin{align} \label{eq_TKE_order}
    \begin{split}
        & U_e\frac{\partial k}{\partial x}+(U-U_e)\frac{\partial k}{\partial x}+V\frac{\partial k}{\partial y}=-\overline{uv}\frac{\partial U}{\partial y}-\frac{\partial \overline{pv}}{\partial y}-\frac{\partial  \overline{\frac{1}{2}(u^2+v^2+w^2)v}}{\partial y}+\nu\frac{\partial^{2}k}{\partial y^{2}}-\epsilon\\
      &\ \frac{u_*^{3}}{\delta} \hspace{1.2cm} \frac{u_*^{3}}{L} \hspace{1.4cm} \frac{u_*^{3}}{L} \hspace{1.3cm} \frac{u_*^{3}}{\delta} \hspace{1cm} \frac{u_*^{3}}{\delta}
      \hspace{1.3cm} \frac{u_*^{3}}{\delta} \hspace{2cm} \nu\frac{u_*}{\delta^{2}}
      \ \ \ \frac{u_*^{3}}{\delta} \\
        &\ O(1) \hspace{.6cm} O(\frac{u_*}{U_e}) \hspace{0.8cm} O(\frac{u_*}{U_e}) \hspace{0.9cm}  O(1) \hspace{0.5cm}  O(1) \hspace{1.3cm} O(1) \hspace{1.1cm} O(Re_*^{-1}) \ \  O(1).
    \end{split}
\end{align}
%The pressure transport and turbulent transpot terms in general are of $O(1).
Note that this order of magnitude analysis is only intended for identifying the leading-order equations. It does not necessarily provide
an accurate estimate of the higher-order terms. An accurate estimate will be made in section 3, after the similarity
variables have been identified.

 From equations \ref{eq_MM_order} -- \ref{eq_TKE_order} we obtain the leading-order mean momentum equation, shear stress budget, and TKE budget.
\begin{equation}
U_e \frac{\partial U}{\partial x}=-\frac{\partial \overline{uv}}{\partial y},
\end{equation}
\begin{equation}
  U_e \frac{\partial  \overline{uv}}{\partial x}=-\overline{u\frac{\partial p}{\partial y}+v\frac{\partial p}{\partial x}}-
  \overline{v^2}\frac{\partial U}{\partial y}.
\end{equation}
\begin{equation}
  U_e \frac{\partial k}{\partial x}=-  \overline{uv}\frac{\partial U}{\partial y}-\frac{\partial \overline{pv}}{\partial y}
  -\frac{\partial  \overline{\frac{1}{2}(u^2+v^2+w^2)v}}{\partial y}-\epsilon.
\end{equation}
We now consider the dilation group of these equation. \textcolor{black}{The velocity-pressure gradient term the shear stress budget scales the same as the production.
The pressure transport and turbulent transport terms in the TKE budget scale the same as the production. Therefore, they should dilate in the same way as the
production terms.} For them to be invariant the exponents must satisfy
\begin{equation}
  g-a=d-b,
\end{equation}
\begin{equation}
  d-a=d_1+g-b,
\end{equation}
and
\begin{equation}
  d_1-a=d+g-b=\frac{3d_1}{2}-b,
\end{equation}
respectively, leading to $d=d_1=2b-2a$ and $g=b-a$, and hence a two-parameter dilation group
\begin{align}
\begin{split}
  \tilde{x}=e^{a}x, \ \tilde{y}=e^bx, \ \tilde{U}-U_e=e^{b-a}(U-U_e), \ \partial \tilde{U}=e^{b-a}\partial U, \\ \widetilde{\overline{uv}}=e^{2b-2a}\overline{uv},
  \ \widetilde{\overline{u^2}}=e^{2b-2a}\overline{u^2}, \
  \widetilde{\overline{v^2}}=e^{2b-2a}\overline{v^2},\ \tilde{u_*}=e^{b-a}u_*. \nonumber
\label{eq_2pgroup}
\end{split}\\
\end{align}

To obtain the one-parameter dilation group under which the outer layer is invariant, an additional relationship is needed.
(If one assumes incorrectly that $U-U_e\sim U_e$, one obtains
$a=b$, the same as \ref{eq_ab}. Therefore, from a mathematical point of view, the problem with this assumption is that it over-specifies
the dilation exponents.) It is obtained by asymptotically matching the outer layer
with the inner layer. One could proceed with \ref{eq_2pgroup} and determine the relationship between $a$ and $b$ when performing matching.
However, this would result in a much lengthier derivation. Here we will instead use \textcolor{black}{an ansatz,}
the logarithmic friction law, to provide this relation. We will show later that the group and the subsequent analysis \textcolor{black}{including matching} indeed lead to the
logarithmic friction law (and also confirms $U-U_e\sim u_*$ as the correct scaling for the velocity defect).
This essentially amounts to guessing the solution of an equation and verifies it later using the equation.
The friction law dilates as
\begin{equation}
 \frac{U_e}{u_*e^g}=\frac{1}{\kappa}\ln\frac{u_*\delta e^{g+b}}{\nu}+C=\frac{1}{\kappa}(\ln\frac{u_*\delta}{\nu}+g+b)+C.
\end{equation}
Since the equation is invariant under the dilation group, we have 
\begin{equation}
 \frac{U_e}{u_*}=e^g\Big\{\frac{1}{\kappa}(\ln\frac{u_*\delta}{\nu}+g+b)+C\Big\}=\frac{1}{\kappa}\ln\frac{u_*\delta}{\nu}+C.
\end{equation}
Therefore
\begin{equation}
 (e^g-1)\Big\{\frac{1}{\kappa}\ln\frac{u_*\delta}{\nu}+C\Big\}=(e^g-1)\frac{U_e}{u_*}=-\frac{g+b}{\kappa}e^g.
\label{eq_eg2}
\end{equation}
This equation provides an implicit relationship between the exponents $g$ and $b$, thereby resulting in a one-parameter dilation
group. 

The relationship between $g$ and $b$ given by equation (\ref{eq_eg2}) is Reynolds-number dependent, and therefore varies with
the downstream location. 
Unlike the Blasius boundary layer, for which the ratios of the exponents in the dilation group are constants (\citealt{Cantwell2002}),
the ratios among $a$, $b$, and $g$ etc., obtained from equations (2.21 and \ref{eq_eg2}) are not, and
depend on $U_e/u_*$ and therefore also depend on $x$, indicating that the group transformation and the symmetry are not only local in the wall-normal
direction, but also in the streamwise direction.
Rather than directly solving the implicit equation \ref{eq_eg2} to determine the relationship between
the exponents of the local dilation group, we examine differential dilation with exponents
$dg$, $da$, and $db$ etc. Equation (\ref{eq_eg2}) now becomes
\begin{equation}
 dg\frac{U_e}{u_*}=-\frac{dg+db}{\kappa}.
\label{eq_dg}
\end{equation}
Thus
\begin{equation}
 dg=-\frac{db}{\kappa\frac{Ue}{u_*}+1}=-\frac{da}{\kappa\frac{Ue}{u_*}+2}.
\label{eq_dg2}
\end{equation}
From the continuity equation, we obtain
\begin{equation}
  dg-da=dc-db, \ dc=2dg.
\end{equation}
The one-parameter local dilation group therefore is
\begin{align} 
\begin{split}
  \tilde{x}=x+dx=e^{da}x, \ \tilde{y}=y+dy=e^{db}y, \ \tilde{U}-U_e=U+dU=e^{dg}(U-U_e),\  & \\ \nonumber
 \tilde{V}=V+dV=e^{2dg}V, \ \widetilde{\overline{uv}}=\overline{uv}+d\overline{uv}=e^{2dg}\overline{uv},
  \ \widetilde{\overline{u^2}}=\overline{u^2}+d\overline{u^2}=e^{2dg}\overline{u^2}, \\
  \widetilde{\overline{v^2}}=\overline{v^2}+d\overline{v^2}=e^{2dg}\overline{v^2},\
  \tilde{u_*}=u_*+du_*=e^{dg}u_*. &
\label{diff_group}
\end{split}\\
\end{align}
It has the infinitesimals
\begin{equation} \label{infinitesimals}
  u_*, \ -(\kappa\frac{U_e}{u_*}+2)x, \ -(\kappa\frac{U_e}{u_*}+1)y, \ (\kappa\frac{U_e}{u_*}+1)\delta,\  2V.
\end{equation}
It  can also be written as
\begin{align}
\begin{split}
  dx=x{da}, \ dy=ydb, \ dU=(U-U_e){dg},\  
 dV=2Vdg, \ d\overline{uv}=2\overline{uv}dg,
  \  \\ \nonumber
  d\overline{u^2}=2\overline{u^2}dg,\ d\overline{v^2}=2\overline{v^2}dg,\
  du_*=u_*dg. &
\label{eq_diff_dilation}
\end{split}\\
\end{align}
Note that the boundary layer thickness $\delta$ dilates in the same way as $y$. From \ref{infinitesimals} or \ref{eq_diff_dilation} we can
obtain the characteristic equations for the group
\begin{equation}
\frac{du_*}{u_*}=-\frac{dy}{y(\kappa\frac{U_e}{u_*}+1)}=-\frac{d\delta}{\delta(\kappa\frac{U_e}{u_*}+1)}=-\frac{dx}{x(\kappa\frac{U_e}{u_*}+2)}=\frac{dV}{2V}.
\label{eq_character}
\end{equation}
From the first and the fourth terms we obtain
\begin{equation} \label{x}
x \sim u_*^{-2}e^{\kappa U_e/u_*},
\end{equation}
and its non-dimensional form
\begin{equation} \label{eq_xa}
\frac{U_ex}{\nu}=Re_x \sim \frac{U_e^2}{u_*^2}e^{\kappa U_e/u_*}.
\end{equation}
Similarly we can obtain
\begin{equation} 
\delta \sim u_*^{-1}e^{\kappa U_e/u_*}, \ V \sim u_*^2.
\end{equation}
The non-dimensional form of $\delta$ is
\begin{equation} \label{deltaa}
\frac{U_e\delta}{\nu}=Re_\delta \sim \frac{U_e}{u_*}e^{\kappa U_e/u_*}.
\end{equation}
Equations \ref{eq_xa} and \ref{deltaa} are functions of $U_e/u_*$, which can be used as a parameter to determine the dependence of $\delta$ on $x$.
To our best knowledge, these are new analytic results that have not been obtained previously.
Taking the ratio of $\delta$ and $x$ we have
\begin{equation} \label{eq_deltax}
\delta /x \sim u_*, \ \ \delta /x \sim u_*/U_e.
\end{equation} 
Non-dimensionalising the variables using \ref{x} - \ref{eq_deltax} and $U_e$, we obtain the similarity variables for the outer layer as
\begin{equation} \label{eq_simvar}
  U_o=\frac{U-U_e}{u_*}, \ V_o=\frac{VU_e}{u_*^2},\ y_o=\frac{yU_e}{xu_*},  \ \overline{uv}_o=\frac{\overline{uv}}{u_*^2},
  \ \overline{u^2_o}=\frac{\overline{u^2}}{u_*^2},   \ \overline{v^2_o}=\frac{\overline{v^2}}{u_*^2}.
\end{equation}
Equation \ref{eq_simvar} defines the full set of similarity variables for the problem for the first time. In particular, the independent variable $y_o$
is defined using the boundary layer parameters ($x$, $U_e$, $u_*$, and $\nu$), similar to that in the Blasius boundary layer.
In previous studies of ZPGTBL, $y_o$ was not fully defined as the normalizing variable was not specified in terms of
the boundary layer parameters, in contrast to that in the Blasius boundary layer.  In addition, $V_o$ was also not fully defined previously.
The dependent variables in \ref{eq_simvar} in general are functions of $y_o$ and a Reynolds number. With the similarity variables fully defined,
we can derive the similarity equations. Since a turbulent boundary layer
is mathematically a singular perturbation problem, we will obtain an approximate solution 
using the method of matched asymptotic expansions without employing a turbulence model.

\subsection{Inner layer symmetries}

We first perform an order-of-magnitude analysis to obtain the leading-order equations for the inner layer. The results for the mean momentum equation are
\begin{align}
\begin{split}      
    &\ U\frac{\partial U}{\partial x}+V\frac{\partial U}{\partial y}=-\frac{\partial \overline {uv}}{\partial y}-\frac{\partial(\overline {u}^{2}-\overline {v}^{2})}{\partial x}+\nu\frac{\partial^{2}U}{\partial y^{2}}\\
    &\ \frac{u_*^{2}}{L} \hspace{1cm} \frac{u_*^{2}}{L} \hspace{1cm} \frac{u_*^{2}}{\delta_\nu} \hspace{0.5cm} \frac{u_*^2\ln(\frac{\delta}{\delta_\nu})}{L} \hspace{0.7cm} \nu\frac{u_*}{\delta_\nu^{2}}\\
    &\   \hspace{0cm} \frac{\delta_\nu}{L} \hspace{1.1cm} \frac{\delta_\nu}{L} \hspace{1.1cm} O(1) \hspace{0.5cm} \frac{\delta_\nu}{L}\ln(\frac{\delta}{\delta_\nu}) \hspace{0.4cm} O(1),\\
   % &\ \frac{\delta_\nu}{\delta}\frac{u_*}{U_e} \hspace{2.1cm} 0(1)  \hspace{0.5cm} \frac{\nu}{U_e\delta}\frac{U_e}{u_*} \hspace{1cm} 0(1)\\
  %  &\ \frac{\nu}{U_e \delta} \hspace{2.4cm} 0(1) \hspace{0.7cm} \frac{\nu}{u_*\delta} \hspace{1.3cm} 0(1)
    \end{split}
\end{align}
\textcolor{black}{where $\delta_\nu$ is the viscous length scale.}
Multiplying the equation by $\delta_\nu/u_*^{2}$ results in the orders of magnitude in the third line. The scaling of the second term on the r.h.s. 
is obtained using the estimate of \cite{Townsend76}.
 %   \begin{align}
  %      \begin{split}
   %     & \frac{u_*}{L} \sim \frac{V}{\delta\nu}, V \sim \frac{\delta \nu}{L}u_*\\
    %    &\ \sim \frac{\delta\nu}{\delta}\frac{\delta u_*}{L}\frac{1}{Re\delta}\frac{u_*}{U_e}u_*\\
     %   &\ \frac{\nu}{x}\sim \frac{\nu}{u_*}\frac{u_*}{L}
      %  \end{split}
   % \end{align}
The orders of magnitude of the shear-stress budget are
    \begin{align} \label{shear_stress_inner_order}
        \begin{split}
          & U\frac{\partial \overline {uv}}{\partial x}+V\frac{\partial \overline{uv}}{\partial y}=-(\overline{ u\frac{\partial p}{\partial y}+v\frac{\partial p}{\partial x})}
          -\frac{\partial \overline{uv^{2}}}{\partial y}-\overline{v^2}\frac{\partial U}{\partial y}+\nu\frac{\partial^{2}\overline{uv}}{\partial y^{2}}-\epsilon_{12}\\
            &\ \frac{u_*^{3}}{L} \hspace{1cm} \frac{u_*^{3}}{L} \hspace{2cm}\frac{u_*^{3}}{\delta_\nu} \hspace{1.3cm}\frac{u_*^{3}}{\delta_\nu} \hspace{1cm} \frac{u_*^{3}}{\delta_\nu} \hspace{0.8cm} \nu\frac{u_*^{2}}{\delta_\nu^{2}}\\
            &\   \hspace{0cm}  \frac{\delta_\nu}{L} \hspace{1cm} \frac{\delta_\nu}{L} \hspace{2cm} O(1)\hspace{1cm}  O(1) \hspace{0.7cm} O(1) \hspace{0.7cm} O(1)\\
         %   &\ Re\delta^{-1} \hspace{6.15cm} 0(1) \hspace{0.75cm} 0(1) \hspace{0.5cm} 0(1)
        \end{split}
    \end{align}
Multiplying the equation by  $\delta_\nu/{u_*^{3}}$ results in the estimates in the third line.  The orders of magnitude of the TKE budget are
\begin{align}\label{TKE_inner_order}
    \begin{split}
      & U\frac{\partial k}{\partial x}+V\frac{\partial k}{\partial y}=-\overline{uv}\frac{\partial U}{\partial y}-\frac{\partial \overline{pv}}{\partial y}
      -\frac{\partial  \overline{\frac{1}{2}(u^2+v^2+w^2)v}}{\partial y}+\nu\frac{\partial^{2}k}{\partial y^{2}}-\epsilon\\
      &\ \frac{u_*^{3}}{L} \hspace{0.8cm} \frac{u_*^{3}}{L} \hspace{1cm} \frac{u_*^{3}}{\delta_\nu} \hspace{1cm} \frac{u_*^{3}}{\delta_\nu}
      \hspace{1.6cm} \frac{u_*^{3}}{\delta_\nu} \hspace{1.7cm}
      \frac{\nu u_*^{2}}{\delta_\nu^{2}} \hspace{0.5cm} \frac{u_*^{3}}{\delta_\nu}\\
        &\ \frac{\delta_\nu}{L} \hspace{0.8cm} \frac{\delta_\nu}{L} \hspace{1cm} O(1) \hspace{.7cm} O(1) \hspace{1.3cm} O(1) \hspace{1.5cm} O(1) \hspace{.3cm} O(1).
    \end{split}
\end{align}
\textcolor{black}{The pressure transport term in principle can of $O(1)$, i.e., it scales as $u_*^3/\delta_\nu$ , although DNS results of \cite{Spalart88} shows that
  the numerical value is small.}

We now perform a Lie group analysis to obtain the similarity variables in the inner layer. The leading-order mean momentum equation is
\begin{equation} \label{momentum_inner}
 0=-\frac{\partial{\overline{uv}}}{\partial{y}}
  + \nu\frac{\partial^2 U}{\partial y^2}.
\end{equation}
The dilation group is
\begin{equation}
 \tilde{y}=e^by,\ \tilde{U}=e^gU, \ \widetilde{\overline{uv}}=e^{2g}\overline{uv}, \ 
\end{equation}
For equation \ref{momentum_inner} to be invariant, the exponents must satisfy
\begin{equation}
  2g-b=g-2b, \ b=-g.
\end{equation}
From the continuity equation, we have
\begin{equation}
  g-a=c-b, \ c=-a.
\end{equation}
\textcolor{black}{These group paameters are also consistent with the dilation properties of \ref{shear_stress_inner_order} and \ref{TKE_inner_order}.} The dilation group now is
\begin{equation}
 \tilde{x}=e^ax,\  \tilde{y}=e^{-g}y,\ \tilde{U}=e^gU, \ \tilde{V}=e^{-a}V, \ \widetilde{\overline{uv}}=e^{2g}\overline{uv}, \ 
\end{equation}
where $a$ and $g$ are related by \ref{eq_eg2}. The infinitesimals of the group are
\begin{equation} \label{infinitesimals}
  u_*, \ -(\kappa\frac{U_e}{u_*}+2)x, \ -y, \ \  (\kappa\frac{U_e}{u_*}+2)V.
\end{equation}
The  characteristic equations for the group are
\begin{equation}
\frac{du_*}{u_*}=-\frac{dy}{y}=-\frac{dx}{x(\kappa\frac{Ue}{u_*}+2)}=\frac{dV}{V(\kappa\frac{U_e}{u_*}+2)}.
\label{eq_character}
\end{equation}
From the first two terms we have
\begin{equation}
 u_*\sim y^{-1}.
\end{equation}
The last two terms lead to
\begin{equation}
 V\sim x^{-1}.
\end{equation}
Therefore we obtain the similarity variables for the inner layer as
\begin{equation} \label{eq_inner_variables}
U_i=\frac{U}{u_*}, \ V_i=\frac{Vx}{\nu}, \ y_i=\frac{yu_*}{\nu}=y^+,  \ \overline{uv_i}=\frac{\overline{uv}}{u_*^2}.
  \ \overline{u^2_i}=\frac{\overline{u^2}}{u_*^2},   \ \overline{v^2_i}=\frac{\overline{v^2}}{u_*^2},
\end{equation}
where $V_i$ is new and has not been properly defined previously.

\section{Approximate solution using matched asymptotic expansions}

\textcolor{black}{In a typical Lie group analysis of differential equations, after the symmetries and similarity varables are identified, the similarity equations are
obtained and the similarity solution is sought. In the case of turbulent flows, the similarity equations are unclosed and cannot be solved without
a turbulence model. Therefore, we employ the method of matched asymptotic expansions to obtain an approximate solution without a turbulence model.
The similarity variables are written as asymptotic expansions with similarity variables identified in section 2 as the leading-order variables. They are
substituted into the (full) ZPGTBL equations to obtain the perturbation equations and to identify the higher-order variables. A higher-order approximate solution are then obtained
using asymptotic matching.}

\subsection{Outer expansions}
We write the outer layer similarity variables as asymptotic expansions
\begin{equation} \label{eq_outer_expansion}
U_o(y_o, Re_*)=U_{o1}(y_o)+\Delta_1(Re_*)U_{o2}(y_o)+\Delta_2(Re_*)U_{o3}(y_o)+\Delta_3(Re_*)U_{o4}(y_o)+ ..., \
\end{equation}
\begin{equation} \label{eq_outer_expansion_uv}
\overline{uv}_o(y_o, Re_*)=\overline{uv}_{o1}(y_o)+\Delta_1(Re_*)\overline{uv}_{o2}(y_o)+\Delta_2(Re_*)\overline{uv}_{o3}(y_o)+\Delta_3(Re_*)\overline{uv}_{o4}(y_o) +..., \
\end{equation}
where $\Delta_k$ are gauge functions, chosen such that the similarity variables $U_{ok}$ etc.~are of order one. Substituting \ref{eq_outer_expansion} into the
mean momentum equation, we obtain the perturbation equation for $U_o$. A detailed derivation is given in Appendix. 
 \begin{equation} \label{U_o_equation}
   \begin{split}
   (-1+\frac{1}{\kappa \frac{U_e}{u_*}+2})y_o\frac{dU_{o1}}{dy_o}-\frac{u_*}{U_e}\frac{U_{o2}}{\kappa \frac{U_e}{u_*}+2}
   +\frac{u_*}{U_e}y_o\frac{dU_{o2}}{dy_o}(-1+\frac{1}{\kappa \frac{U_e}{u_*}+2}) \\
   -Re_*^{-1}U_{o3}\big(1+\frac{-2}{\frac{\kappa U_e}{u_*}+2}\big)+Re_*^{-1}\frac{dU_{o_3}}{dy_o}\big(-y_o+\frac{y_o}{\frac{kU_e}{u_*}+2}\big)\\
   -2\frac{u_*^3}{U_e^3}U_{o_4}\big(1+\frac{-2}{\frac{kU_e}{u_x}+2}\big)+\frac{u_*^2}{U_e^2}\frac{dU_{o_4}}{dy_o}\big(-y_o+\frac{y_o}{\frac{kU_e}{u_x}+2}\big) \\
   -\frac{u_*}{U_e}\frac{U_{o1}^2}{\kappa\frac{U_e}{u_*}+2} + \frac{u_*}{U_e}U_{o1}\Big\{\frac{dU_{o1}}{dy_o}(-y_o+\frac{y_o}{k\frac{u_e}{u_*}+2})
     -\frac{u_*}{U_e}\frac{U_{o2}}{\kappa \frac{U_e}{u_*}+2}    +\frac{u_*}{U_e} \frac{dU_{o2}}{dy_o}(-y_o+\frac{y_o}{k\frac{u_e}{u_*}+2})\Big\}\\
     +\frac{u_*}{U_e}V_o\frac{dU_o}{dy_o}\\
   =-\frac{d\overline{uv_o}}{dy_o}+2\frac{u_*}{U_e}\frac{(\overline{u_o^{2}}-\overline{v_o^{2}})}{k\frac{U_e}{u_*}+2}
     -\frac{u_*}{U_e}\big(-y_o+\frac{1}{k\frac{U_e}{u_*}+2}\big)\frac{d(\overline{u_{o1}^{2}}-\overline{v_{o1}^{2}})}{dy_o}-\frac{U_e^2}{u_*^2}\frac{(\overline{u_{o2}^{2}}-\overline{v_{o2}^{2}})}{k\frac{U_e}{u_*}+2}\\
        -\frac{u_*^2}{U_e^2}\big(-y_o+\frac{1}{k\frac{U_e}{u_*}+2}\big)\frac{d(\overline{u_{o2}^{2}}-\overline{v_{o2}^{2}})}{dy_o}+Re_*^{-1}\frac{\partial^{2}U_o}{\partial y_o^2}.
   \end{split}
   \end{equation}
The leading-order mean momentum equation is
\begin{equation} \label{leading_outer_momentum}
  -y_o\frac{dU_{o1}}{dy_o}=-\frac{d\overline{uv}_{o1}}{dy_o},
\end{equation}
which is identical to that obtained by \cite{TL72}, suggesting that their definition of the boundary layer thickness and the non-dimensional wall-normal coordinate
are equivalent to $\delta$ and $y_o$ in the present work at the leading order. However, their definition would not lead to the higher-order
similarity equations derived here as they will involve more integral variables. The second-order and third-order equations can be obtained by
collecting the terms containing $Re_*$ and $u_*^2/U_e^2$ respectively in \ref{U_o_equation}. 

Similarly, the perturbation equation for the Reynolds shear stress is derived in Appendix.
\begin{equation} \label{shear_budget}
  \begin{split}
    -\frac{2\overline {uv}_o}{k\frac{U_o}{u_*}+2}
  +\frac{d\overline{uv}_{o1}}{dy_o}y_o(-1+\frac{1}{{k\frac{U_e}{u_*}+2}})- \frac{u_*}{U_e}\frac{\overline{uv}_{o2}}{k\frac{U_o}{u_*}+1}
  +\frac{u_*}{U_e}\frac{d\overline{uv}_{02}}{dy_o}(y_o+\frac{y_o}{k\frac{U_o}{u_*}+2})\\
  -Re_*^{-1}\overline{uv}_{o3}\big(1+\frac{-2}{\frac{\kappa U_e}{u_*}+2}\big)+Re_*^{-1}\frac{d\overline{uv}_{o3}}{dy_o}\big(-y_o+\frac{y_o}{\frac{kU_e}{u_*}+2}\big)\\
  -2\frac{u_*^3}{U_e^3}\overline{uv}_{o4}\big(1+\frac{-2}{\frac{kU_e}{u_x}+2}\big)+\frac{u_*^2}{U_e^2}\frac{d\overline{uv}_{o4}}{dy_o}\big(-y_o+\frac{y_o}{\frac{kU_e}{u_x}+2}\big) \\
  -\frac{u_*}{U_e}\frac{2U_o\overline{uv_o}}{k\frac{U_o}{u_*}+2}
       +\frac{u_*}{U_e}U_o\frac{d\overline{uv}_{o2}}{dy_o}y_o(-1+\frac{1}{\frac{kU_e}{u_x}+2})+(\frac{u_*}{u_e})^{2}U_{o_1}\frac{-2\overline{uv}_{o2}}{\frac{kU_e}{u_x}+2}\\
       -\frac{u_*^2}{U_e^2}U_{o1}\frac{d\overline{uv}_{o2}}{dy_o}(-1+\frac{-1}{\frac{kU_e}{u_x}+2})+\frac{u_*}{U_e}V_o\frac{\partial \overline{uv_o}}{\partial y_o}\\
       =-\overline{\Big(u\frac{\partial p}{\partial y}+v\frac{\partial p}{\partial x}\Big)}_o -\overline{v_o^{2}}\frac{\partial U_o}{\partial y_o}+Re_*^{-1}\frac{\partial^{2}\overline{uv}_o}{dy_o^{2}}.\\
     \end{split}
 \end{equation}
The leading-order shear-stress budget is
\begin{equation}
  -y_o\frac{d\overline{uv}_{o1}}{dy_o}=-\overline{\Big(u\frac{\partial p}{\partial y}+v\frac{\partial p}{\partial x}\Big)}_o
  -\overline{v_{o}^2}\frac{dU_{o1}}{dy_o}.
  \end{equation}
The mean momentum equation and the Reynolds-stress budget are not a closed set of equations. Solving them directly requires modeling the unclosed
velocity-pressure interaction term. However,
using the method of matched asymptotic expansions, we can obtain an approximate solution in the matching layer \textcolor{black}{without a turbulence model}. 
Since the outer expansions are not valid in the near wall region (the inner layer), inner expansions are needed
to approximate the solution in this region, and are obtained in the following. 

\subsection{Inner expansions}

Similar to the outer similarity variables, the inner similarity variables depend on $y^+$ and $Re_*$. We write them as asymptotic expansions,
\begin{equation} \label{eq_inner_expansion}
U_i(y^+, Re_*)=U_{i1}(y^+)+\delta_1(Re_*)U_{i2}(y^+)+\delta_2(Re_*)U_{i3}(y^+)+ ..., \
\end{equation}
\begin{equation} \label{eq_outer_expansion_uv}
\overline{uv}_i(y^+, Re_*)=\overline{uv}_{i1}(y^+)+\delta_1(Re_*)\overline{uv}_{i2}(y^+)+\delta_2(Re_*)\overline{uv}_{i3}(y^+)+\delta_3(Re_*)\overline{uv}_{i4}(y^+)+ ..., \
\end{equation}
where $\delta_k$ etc.~are gauge functions. Substituting \ref{eq_inner_variables}, we obtain  the inner equations. The details are given in Appendix.
The inner mean momentum equation up to the order of $Re_\delta^{-1}u_*/U_e$ is
\begin{equation}
     \begin{split}             
       -Re_{\delta}^{-1}\frac{u_*}{U_e}\Big\{U_i^{2}+y^{+}\frac{dU_i}{dy^{+}}\Big\}+Re_\delta^{-1}V_i \frac{dU_i}{dy^{+}}
       =-\frac{\partial \overline {uv_i}}{\partial y^{+}}+\frac{d^{2}U_i}{dy^{+2}}\\
       -Re_\delta^{-1}\overline{u_i^{2}}-Re_\delta^{-1}y^{+}\frac{d \overline{u_i^{2}}}{d y^{+}}+
       2Re_\delta^{-1}\frac{u_*}{U_e}\overline{v_i^{2}}+Re_\delta^{-1}\frac{u_*}{U_e}y^{+}\frac{\partial \overline{v_i^{2}}}{\partial y^{+}}.
     \end{split}
 \end{equation}
The inner shear-stress budget up to the order of $Re_\delta^{-1}u_*/ U_e$ is
\begin{equation}
     \begin{split}      
       -Re_\delta^{-1}U_i\big(\frac{\overline{uv_i}}{k\frac{U_e}{u_*}+2}+y^{+}\frac{d \overline{uv_i}}{d y^{+}}\big)+Re_\delta^{-1}V_i\frac{d \overline{uv_i}}{d y^{+}}
       =-\overline{\Big(u\frac{\partial p}{\partial y}+v\frac{\partial p}{\partial x}\Big)}_i \\
       -\overline {v_i^{2}}\Big(\frac{d U_{i1}}{d y^{+}}+Re_\delta^{-1}\frac{d U_{i2}}{d y^{+}}+Re_\delta^{-1}\frac{u_*}{\kappa U_e}\frac{d U_{i3}}{d y^{+}}\Big)
       +\frac{d^{2}\overline{uv}_{i1}}{d y^{+2}}+Re_\delta^{-1}\frac{d^{2}\overline{uv}_{i2}}{d y^{+2}}
       +Re_\delta^{-1}\frac{u_*}{\kappa U_e}\frac{d^{2}\overline{uv}_{i3}}{d y^{+2}}.
     \end{split}
 \end{equation}

%The leading-order equations are

\subsection{Matching the expansions}

We now match the velocity expressed as the outer and inner expansions 
\begin{equation} \label{eq_expansion_outerv}
U=U_e+u_*U_o=U_e+u_*\Big\{U_{o1}(y_o)+\frac{u_*}{U_e}U_{o2}(y_o)+Re_*^{-1}U_{o3}(y_o)+\frac{u_*^2}{U_e^2}U_{o4}(y_o)+ ...\Big\}, \
\end{equation}
\begin{equation} \label{eq_expansion_innerv}
U=u_*U_i=u_*\Big\{U_{i1}(y^+)+Re_\delta^{-1}U_{i2}(y_o)+Re_\delta^{-1}\frac{u_*}{U_e}U_{i3}(y_o)+ ...\Big\}. \
\end{equation}
Asymptotically matching the leading-order terms $U_e+u_*U_{o1}$ and $u_*U_{i1}$ results in the log law
\begin{equation}
  U_{o11}=\frac{1}{\kappa}\ln y_o +C, \ \ U_{i11}=\frac{1}{\kappa}\ln y^+ +B.
\end{equation}
Matching $(u_*/U_e)U_{o2}$ and $Re_\delta^{-1}U_{i2}$ results in
\begin{equation}
  U_{o21}\sim y_o, \ \ U_{i2} \sim y^+.
\end{equation}
Matching $Re_*^{-1}U_{o3}$  and $U_{i1}$ gives
\begin{equation}
  U_{o3}\sim y_o^{-1}, \ \ U_{i12} \sim (y^+)^{-1}.
\end{equation}
Matching the outer expansion with $U_{i3}$ requires the so-called block matching involving several terms in the outer expansion (\citealt{BO1978}),
\begin{equation}
  U_{o12}+\frac{u_*}{U_e}U_{o22}+\frac{u_*^2}{U_e^2}U_{o4}.
\end{equation}
The results are
\begin{equation}
  U_{o12}\sim y_o, \ \ U_{o22}\sim y_o\ln y_o, \ \ U_{o4}\sim y_o\ln^2 y_o,\nonumber
\end{equation}
\begin{equation}
  U_{i3}\sim y^+\ln^2 y^+.
\end{equation}
Inserting the matching results into \ref{eq_expansion_outerv} and \ref{eq_expansion_innerv} we obtain the outer expansion as
\begin{equation} \label{eq_matchedUo}
U=U_e+u_*\Big\{\frac{1}{\kappa}\ln y_o+C+A_uy_o+B_u\frac{u_*}{U_e}y_o+A_u\frac{u_*}{U_e}y_o\ln y_o+C_uRe_*^{-1}y_o^{-1}+ A_u\frac{u_*^2}{U_e^2}y_o\ln ^2y_o+ ...\Big\},
\end{equation}
and the inner expansion as
\begin{equation} \label{eq_matchedUi}
U=u_*\Big\{\frac{1}{\kappa}\ln y^++B+C_u(y^+)^{-1} +B_u Re_\delta^{-1}y^++ A_uRe_\delta^{-1}\frac{u_*}{U_e}y^+\ln ^2y^++ ...\Big\}.
\end{equation}

From \ref{eq_matchedUo} and \ref{eq_matchedUi} we obtain the friction law
\begin{equation} \label{friction_law}
\frac{U_e}{u_*}=\frac{1}{\kappa}\ln \frac{y^+}{y_o}+B-C=\frac{1}{\kappa}\ln \frac{u_*^2x}{U_e\nu}+B-C,
\end{equation}
\textcolor{black}{which also confirms the anstaz used in section 2.1 to obtain the outer layer symmetry.}
Note that when obtaining the friction law, each term in the higher-order expansions in the outer and inner layers match and cancel each other except for any log-law-like
terms, which do not cancel, just like the (leading-order) log law. However, log-law-like terms come from matching terms that have the same scales
in both the outer and inner expansions (\citealt{TD19b}). From the gauge functions it is clear that only the leading-order terms have the same scale.
Therefore, there are no higher-order low-law-like terms; therefore the logarithmic friction law \ref{friction_law} is accurate.

Similarly we obtain the matching results for $\overline{uv}$ as
\begin{equation} \label{eq_matched_uvo}
\overline{uv}=u_*^2\Big\{-1+A_{uv}y_o+B_{uv}\frac{u_*}{U_e}y_o+A_{uv}\frac{u_*}{U_e}y_o\ln y_o+C_{uv}Re_*^{-1}y_o^{-1}+ A_{uv}\frac{u_*^2}{U_e^2}y_o\ln ^2y_o+ ...\Big\},
\end{equation}
and
\begin{equation} \label{eq_matched_uvi}
\overline{uv}=u_*^2\Big\{-1+C_{uv}(y^+)^{-1} +B_{uv} Re_\delta^{-1}y^++ A_{uv}Re_\delta^{-1}\frac{u_*}{U_e}y^+\ln ^2y^++ ...\Big\}.
\end{equation}

From the leading-order momentum equation \ref{leading_outer_momentum}, we obtain $A_{uv}=1/\kappa$.
While the dilation group \ref{diff_group} is Reynolds number dependent, the leading-order expansions (solution) of the boundary layer
equations are not.

\subsection{Preliminary comparison with measurements}

Determining the non-dimensional coefficients, including the expansion coefficients in the theoretical prediction requires comparisons with experimental results.
In particular, given the large amount of existing data, determining the expansion coefficients require extensive comparisons, and
is beyond the scope of the present work, which focuses on the theoretical development. Such comparisons will be
carried out in future works, perhaps in collaboration with experimental groups.
In this work we  make  preliminary comparisons of the prediction of the non-dimensional velocity $U_e/u_*$ and
the non-dimensional outer layer thickness $Re_\delta =U_e\delta_{99}/\nu$ as a function of the non-dimensional downstream distance $Re_x=U_ex/\nu$ with
the experimental data of \cite{Marusic2015} (the SP40 configuration). The measured values of $U_e/u_*$ are used as the parameter to obtain the theoretical values of
$Re_x$ and $Re_\delta$ using equations \ref{eq_x} and \ref{delta}. The  von K\'{a}rm\'{a}n
constant $\kappa=0.420$ and the non-dimensional coefficient for $\delta_{99}$ are determined by
fitting equation \ref{delta} to the experimental data. The virtual origin of $x=-1.744$ m and the non-dimensional coefficient for $x$ are obtained
by fitting \ref{eq_x} to the data. In particular, the values of $U_e/u_*$ and $\delta_{99}$ at $x=1.6 $ m are used to determine the non-dimensional coefficients.
The kinematic viscosity is taken as the value in \cite{Marusic2015}, $\nu=15.1\times 10^{-6}$ m/s$^2$.
The results are
\begin{equation} \label{eq_x}
Re_x= 0.06024\frac{U_e^2}{u_*^2}e^{\kappa U_e/u_*},
\end{equation}
and
\begin{equation} \label{delta}
Re_\delta =0.02204 \frac{U_e}{u_*}e^{\kappa U_e/u_*}.
\end{equation}
We then have
\begin{equation} \label{}
\delta_{99}=0.3659\frac{xu_*}{U_e}, \ \ y_o=\frac{yU_e}{xu_*}=0.3659\frac{y}{\delta_{99}}.
\end{equation}
\begin{figure}
\centering
\includegraphics[width=3.6in,height=2.4in]{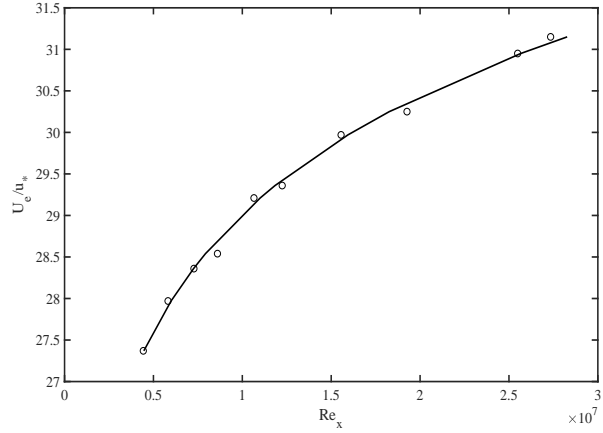}
\caption{Non-dimensional velocity $U_e/u_*$ as a function of the non-dimensional downstream distance $Re_x=U_e x/\nu$. Circles: experimental data
from Marusic (2015) (the SP40 configuration); Solid line: Theoretical prediction of equation \ref{eq_x}.}
\vspace{0.5in}
\label{fig1}
\end{figure}

\begin{figure}
\centering
\includegraphics[width=3.6in,height=2.4in]{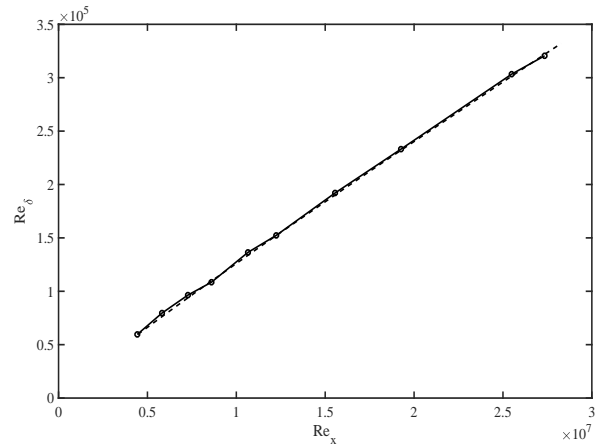}
\caption{Non-dimensional boundary layer thickness $U_e \delta/\nu$ as a function of the non-dimensional downstream distance $Re_x=U_e x/\nu$. Circles: experimental data
from Marusic (2015) (the SP40 configuration); Dashed line: Theoretical prediction of equation \ref{delta}.}
\label{fig2}
\end{figure}

Figures 1 and 2 show that with these numerical values of the coefficients, the theoretical prediction has an excellent agreement with the experimental results.
It is interesting to note that the values of the von  K\'{a}rm\'{a}n constant
obtained is different from the value of $0.384$ obtained in the same experiment and  by \cite{Nagib2007},
but much closer to that of \cite{Vallikivi2015a} (0.40) and the typical value of 0.421
in pipe flows (\citealt{McKeon2004,McKeon2007}). We emphasize that these are preliminary comparisons with a single set of experimental data.
Nevertheless, they might still shed some light on the way in which the von  K\'{a}rm\'{a}n constant
is determined from the velocity profile and needs further attention.

\textcolor{black}{Here we also provide a list of predictions to be evaluated using experimental data in future works.}

\begin{enumerate}[(1)]

\item \textcolor{black}{Comprehensive assessment of the downstream evolution of the friction velcity and the boundary layer thickness. In particular, the virtual origion, the
  von  K\'{a}rm\'{a}n constant and the non-dimensional coefficients in \ref{eq_x} and \ref{delta}  should be evaluated.}

\item  \textcolor{black}{The logarithmic friction law \ref{friction_law}. Note that there are no higher-order corrections to this law. Therefore, when plotting
  $U_e/u_*$ vs $\ln(u_*^2x/U_e\nu)$, a straightlie should extend down to relatively modest Reynolds numbers. Such comparisons will also provide another evaluation of
the  von  K\'{a}rm\'{a}n constant}

\item \textcolor{black}{Comprehensive evaluation of the expansion coefficients in the asymptotic expansions \ref{eq_matchedUo} or \ref{eq_matchedUi}
of  velocity profile in the overlapping region. The extent of the log region can be evaluated and  comparisons with channel flows can be made.}

\item  \textcolor{black}{A new evaluation of the  von  K\'{a}rm\'{a}n constant using the measured mean velocity profile and the predicted higher-order prediction in the overlapping
  region (log law plus the higher-order corrections), rather than the log law alone, as done previously. This allows determination of the constant using data obtained
at moderate Reynolds numbers, rather than chasing the seemingly ellusive high Reynolds numbers.}

 \item   \textcolor{black}{Comprehensive evaluation of the expansion coefficients in the asymptotic expansions \ref{eq_matched_uvo} or \ref{eq_matched_uvi}
 of Reynolds shear-stress profile in the overlapping region.}
 
  \end{enumerate}

\section{Discussions}

The asymptotic expansions are an approximate solution of the boundary layer equations in the matching layer. They allow us to
compare and contrast ZPGTBL with channel flows.
In channel flows (and pipe flows), the leading-order mean momentum equation in the outer layer is a balance between the shear stress derivative and
the mean pressure gradient (\citealt{Afzal1976}), with the latter imposing the linear variation of the leading-order (linear) variation of the Reynolds shear
stress. The Reynolds shear-stress budget is a balance between shear production and velocity-gradient--pressure
interaction.  While the mean velocity itself does not appear
in the mean momentum equation, the mean velocity gradient is ``adjusted'' to balance the Reynolds stress budget. This is where the constant stress
condition is needed for the log law (through the pressure-strain-rate correlation). 

In ZPGTBL the leading-order mean momentum balance is between the mean advection due to the free-stream velocity $U_e$ and the shear stress derivative.
The leading-order (linear) variation of the Reynolds shear stress results from the log law.
It is interesting to observe that while the non-dimensional leading-order Reynolds shear stress derivative equals
$1/\kappa$, the leading-order shear stress in the boundary layer can be written as
\begin{equation}
  \overline{uv}=u_*^2\Big\{-1+\frac{1}{\kappa}y_o\Big\}=u_*^2\Big\{-1+\frac{0.3659}{\kappa}\frac{y}{\delta_{99}}\Big\}.
\end{equation}
The coefficient for the second term on the r.h.s. is not very different from the value of one in channel flows.
The mean advection depends on the growth of the boundary layer thickness and
the variations of $u_*$; therefore the mean momentum balance is more delicate.
The Reynolds shear stress balance is among the mean advection, production and velocity gradient-pressure interaction respectively.
Again, due to the mean advection, the balance is also more delicate.
However, in the near-wall region ($y_o \ll 1$), the advection term is of higher-order.
Therefore the Reynolds shear stress balance is similar to that in channel flows.
%Although the velocity gradient appears in both the mean momentum equation and the shear stress budge, in the former the shear stress appears in
%its derivative, therefore the mean momentum equation does not interfer with the constant stress condition. Rather, it might only impose a condition on

The functional forms of the leading-order terms for both the outer ($U_o=(1/\kappa)\ln y_o+C+A_u y_o$) and inner expansions
($U_i=(1/\kappa)\ln y^++B+C_u (y^+)^{-1}$) are identical for channel flows and ZPGTBL. 
Therefore the von K\'{a}rm\'{a}n constant and the extent of the log layer at the leading order should be asymptotically identical for both flows.
Physically this is because the leading-order scaling of the turbulence fluctuations (e.g., the velocity variances and
the pressure-strain-rate correlation) are the essentially same
for both channel flows and the boundary layer. Therefore, there is no reason to believe that the log law would not be universal. 
However the values of $A_u$ and $C_u$ determine the onset of the log layer, i.e., the lowest $y^+$ and the highest $y/\delta$ values
for the log layer. Given that the leading-order Reynolds shear-stress budget contains an advection term, we might expect
$A_u$ to be different from its counterpart in channel flows. 

In channel flows, the lowest (second-order) order correction terms to (or deviations from) the leading-order terms are of order $Re_{*}^{-1}$ in both
the outer and inner layers, whereas in ZPGTBL they are of order $u_*/U_e \sim \ln^{-1}Re_{*}$ in the outer layer, larger
than in channel flows, and are of order $Re_{\delta}^{-1}$ in the inner layer, smaller than in channel flows.
This suggests that the deviations from the leading order
are larger toward the outer layer and smaller toward the inner layer in ZPGTBL than in channel flows.

In channel flows there is a second-order (and more at the higher orders) logarithmic term in the mean velocity  (\citealt{Afzal1976}),
indicating that the observed coefficient
for the logarithmic part of the profile, which is usually taken as the (apparent) von K\'{a}rm\'{a}n constant, is Reynolds-number dependent.
Note that the von K\'{a}rm\'{a}n constant is defined as the coefficient for the leading-order logarithmic profile.
Such a term is absent in ZPGTBL. As a result, at finite Reynolds numbers
we will observe departures from the log law in ZPGTBL, rather than a Reynolds-number-dependent (apparent) von K\'{a}rm\'{a}n constant. 

The physics related to the corrections are also different in the two flows. In channel flows it is due to (higher-order) viscous effects in the outer layer
and (higher-order) variations of the shear stress in the inner layer respectively. The former depends explicitly on the Reynolds number whereas the latter has
an implicit dependence through other variables (variations of the shear stress).
In ZPGTBL, the leading-order the corrections in the outer layer are caused by the advection due to the velocity defect,
advection due to the wall-normal mean velocity, and the
streamwise derivative of the velocity variance differences. The dependence of the Reynolds number is implicit,  through the effects of
the Reynolds number on the growth of the boundary layer thickness and the variations of $u_*$, as none of the variables
are directly dependent on the viscosity. In the inner layer, the corrections are caused by advection due to the vertical velocity. 

The leading-order logarithmic friction law for channel flows is accurate to $O(Re_*^{-1})$. For ZPGTBL, equation \ref{friction_law} indicates that
it is accurate, without any higher-order corrections.
%to at least $O(u_*^3/U_e^3)$. However, it is quite possible that all the higher-order term in the outer expansion have  equal terms in the
%inner expansion. They will cancel each other when obtaining the friction law, until (but unlikely) a higher-order log law appears as in channel flows.
Therefore, the logarithmic friction law  is more accurate in ZPGTBL than in channel flows.

%It is often stated in the literature that the true log law will not be observed  until certain Reynolds numbers are reached.
%This may be a misunderstanding of the nature of the log law.

\section{Conclusions}

We performed a symmetry analysis of the equations for ZPGTBL using Lie dilation groups, and obtained local, leading-order symmetries of
the equations. The full set of similarity variables was obtained from the characteristic equations of the group.
The dependent similarity variables, which are generally functions of the independent similarity variable (the local non-dimensional normal
coordinate) and Reynolds numbers, were written as asymptotic expansions, with the gauge functions depending on the Reynolds numbers.
Using the asymptotic expansions the perturbation equations for the outer
and inner layers were obtained and the gauge functions were determined. Matching the expansions resulted in an approximate similarity solution of
ZPGTBL in the overlapping layer. Expansion terms up to the third order were obtained.
To our best knowledge, these results have not been obtained previously.
Furthermore, they were obtained from first principles without any major assumptions.
%The fundamental assumption on which the theory is based is the
%finiteness of the TKE dissipation as viscosity approaches zero (but does not equal zero), which ensures Taylor's scaling of the dissipation
%and the invariance of the non-dimensional dissipation.
The main results are as follows.
%This assumption leads to the Reynolds number
%invariance of the spectral transfer at large scales and hence of the energy-containing statistics at high Reynolds numbers.

\begin{enumerate}[(1)]%[label=\arabic*)]

\item We showed that similar to the Navier-Stokes equations at high Reynolds numbers (turbulent flows), the Reynolds-averaged boundary layer
  equations for ZPGTBL do not have global symmetries, only have local, approximate symmetries. We obtained the leading-order Lie dilation symmetries,
 i.e., invariance under Lie dilation groups, in both the outer and inner layers.
  
\item We showed that the friction $u_*$ is the scale for the velocity defect. Using the free-stream $U_e$ as the scale, as suggested by some
previous studies, would lead to an inconsistency with the scaling of the TKE dissipation rate.

\item We derived analytically from the boundary layer equations the non-dimensional friction velocity $u_*/U_e$ and the non-dimensional boundary layer thickness
$U_e\delta/\nu$ as  functions
of the non-dimensional downstream distance $U_ex/\nu$ using
the characteristic equations of the group. When using the von K\'{a}rm\'{a}n constant, the virtual origin of the boundary layer, and two non-dimensional
coefficients determined from the experimental data of \cite{Marusic2015},
the theoretical prediction shows excellent agreement with the data.

\item From the symmetry analysis, the independent similarity variable for the outer layer wall-normal coordinate was obtained using the boundary layer
  parameters as $y_o=yU_e/(xu_*)$, analogous to that in the Blasius boundary layer. Previously the
non-dimensionalising variable was not fully defined, i.e., not using the boundary layer parameters.

\item The leading-order similarity equations, which are ordinary differential equations, were obtained.
The dependent similarity variables, written as asymptotic expansions, were used to derive the perturbation equations, which also contain the
higher-order similarity equations and have not been obtained previously.
The gauge functions for the outer layer are determined as $u_*/U_e$, $Re_*^{-1}$, $u_*^2/U_e^2$, ... Those for the inner layer
are $Re_\delta^{-1}$, $Re_\delta^{-1} u_*/U_e$, ... These gauge functions are in contrast to those for channel flows, which are $Re_*^{-1}$, $Re_*^{-2}$, ..., for
both the outer and inner layers.
 The approximate solution in the overlapping layer as asymptotic expansions were obtained. Using the method of
matched asymptotic expansions to obtain an approximate solution avoided a turbulence model.

\item While the leading-order outer equations for ZPGTBL
  are different from those for channel flows, the leading-order expansion terms in the overlapping layer are formally identical.
  This result provides analytic support to the notion that the log law and the broader
  leading-order mean flow are asymptotically universal in this layer.

\item The higher-order expansions for the ZPGTBL and channel flows are different.
With the log law at the leading order in ZPGTBL without any other logarithmic terms, the logarithmic friction law is accurate. By contrast, in channel
flows it has a logarithmic term at the second order. Therefore the friction law is only accurate at the leading order.

\item The second-order corrections to the leading-order outer-layer expansions are asymptotically larger
in ZPGTBL than in channel flows, whereas those to the leading-order inner-layer expansions are smaller in ZPGTBL than in channel flows.
Due to these differences, we might expect the onset and possibly the extent of the logarithmic profile in measurements and simulations 
 to be different. This issue will be investigated in a future work.

\end{enumerate}

The asymptotic expansions obtained in the present work will also provide a more systematic approach to assess the asymptotic behaviours of ZPGTBL
using experimental data. For example,
the log law is often investigated using the measured mean velocity profile.
While the mean profile exhibits an approximate logarithmic dependence only at sufficiently high Reynolds numbers,
the log law is a leading-order term in the profile, a result of matching the leading-order outer and inner profiles with the same
velocity scale (\citealt{TD19b}). It is present even at moderate Reynolds numbers. At
such Reynolds numbers it may be obscured by the other leading-order and higher-order terms in the mean velocity profile. However, in principle it
can be obtained experimentally by determining the coefficients of the other terms  in \ref{eq_matchedUo}, e.g., using
measurements at several Reynolds numbers and then subtracting these terms from the mean profile. The issue of the asymptotic scaling of ZPGTBL
perhaps could be better addressed through collaboration among the theoretical and experimental groups.

Declaration of Interests. The authors report no conflict of interest.

{\bf Acknowledgments}

The author thanks Professors Zellman Warhaft and Katepalli Sreenivasan  for reading the manuscript and for providing valuable comments.
He is also thankful to Professor Sreenivasan for encouraging him to look into issues in classical engineering turbulent boundary layers.
%when inviting him for a seminar at New York University on the atmospheric boundary layer.
This work was supported by
the National Science Foundation under grant AGS-2054983.

\appendix
\section{Derivation of the outer and inner layer perturbation equations}
%\chapter{Appendices}
%\markboth{Appendices}{}
%\addcontentsline{toc}{chapter}{Appendices}
\renewcommand{\thesection}{A} %\arabic{section}}

%\begin{equation} \label{eq}
 %  U=u_*\Big\{U_{o_1}(y_o)+\Delta_1(Re_*)U_{o_2}(y_o)+\Delta_2(Re_*)U_{o_3}+\Delta_3(Re_*)U_{o_2}(y_o)\Big\}
%\end{equation}
Equation \ref{eq_MM_order} suggests that $\Delta_1=u_*/U_e\approx \ln^{-1} Re_*$, which is confirmed by \ref{defect_advection}. It also shows that
 $\Delta_2=Re_*^{-1}$. With these gauge functions the perturbation and similarity
equations for the mean momentum equation in the outer layer are derived as follows.
\begin{equation}
\frac{\partial U}{\partial x}=\frac{du_*}{dx}U_{o}+u_*\frac{d}{dx}\Big\{U_{o_1}+\Delta_1U_{o_2}+Re_*^{-1}U_{o_3}+\Delta_3(Re_*)U_{o_2}(y_o)\Big\},
\end{equation}
where
\begin{equation}
\begin{split}
  \frac{d}{dx}\Big\{U_{o_1}+\Delta_1U_{o_2}\Big\}=\frac{dU_{o_1}}{dy_o}yU_e(\frac{-1}{x^2u_*}+\frac{-1}{xu_*}\frac{du_*}{dx}) +\frac{d\Delta_1}{dx}U_{o_2}+\Delta_1\frac{dU_{o_2}}{dy_o}yu_e  \\ 
  =\frac{dU_{o1}}{dy_o}(\frac{-y_o}{x}+\frac{-y_o}{u_*}\frac{du_*}{dx})+ \frac{1}{U_e}\frac{du_*}{dx}U_{o2}
   +\Delta_1\frac{dU_{o2}}{dy_o}(\frac{-y_o}{x}+\frac{-y_o}{u_*}\frac{du_*}{d_x}).
 \end{split}
\end{equation}
The non-dimensional form of its contribution to the advection is
\begin{align*}
\begin{split}
\frac{x}{U_eu_*}U_eu_*\frac{d}{dx}\Big\{U_{o_1}+\Delta_1U_{o_2}\Big\}=\frac{dU_{o_1}}{dy_o}(-y_o-\frac{x}{u_*}y_o\frac{du_*}{dx})
+\frac{x}{U_e}\frac{du_*}{dx}U_{o2}+\Delta_1\frac{dU_{o2}}{dy_o}(y_o+\frac{x}{u_*}y_o\frac{du_*}{dx})
\end{split}
\end{align*}
\begin{equation}
\begin{split} \label{advection1}
=\frac{dU_{o1}}{dy_o}(-y_o+\frac{y_o}{\kappa \frac{U_e}{u_*}+2})-\frac{u_*}{U_e}\frac{U_{o2}}{\kappa \frac{U_e}{u_*}+2}
 +\Delta_1\frac{dU_{o2}}{dy_o}(-y_o+\frac{y_o}{\kappa \frac{U_e}{u_*}+2}).
 \end{split}
 \end{equation}
The term
\begin{equation}
 \frac{1}{\kappa \frac{U_e}{u_*}+2}=\frac{u_*}{\kappa U_e}-2\Big(\frac{u_*}{\kappa U_e}\Big)^2+...,
\end{equation}
and therefore contains both $\Delta_1$ and $\Delta_3$. The $U_{o3}$ term is
 \begin{equation} \label{advection2}
   \begin{split}
       \frac{d}{dx}(Re_*^{-1}U_{o3})=\frac{dRe_*^{-1}}{dx} U_{o3}+Re_*^{-1}\frac{dU_{o3}}{dx}\\ 
       =-Re_*^{-2}\frac{dRe_*}{dx}+Re_*^{-1}\frac{dU_{o3}}{dy_o}yU_e(\frac{-1}{x^2u_*}+\frac{-1}{xu_*^2}\frac{du_*}{dx})\\ 
       =-Re_*^{-2}\frac{U_{o_3}}{\nu}\frac{1}{U_e}(u^2_*+2xu_*\frac{du_*}{dx})+Re_*^{-1}\frac{dU_{o3}}{dy_o}yU_e(\frac{-1}{x^2u_*}+\frac{-1}{xu_*^2}\frac{du_*}{dx}).
       \end{split}
   \end{equation}
   Its non-dimensional form is
   \begin{equation*}
       \begin{split}
           \frac{x}{U_o u_*}U_eu_*\frac{dRe_x^{-1}U_{o3}}{dx}=-\frac{u_*xRe_*^{-2}}{u_*}\frac{U_{o3}}{\nu}(u_*^2+2xu_*\frac{du_*}{dx})+\\
           \frac{U_exu_*}{U_e u_*}Re_*^{-1}\frac{dU_{o3}}{dy_o}yU_e(\frac{-1}{x^2u_*}+\frac{-1}{xu_*^2}\frac{du_*}{dx})-\\
           =-Re_*^{-2}U_{o3}\big(\frac{u_*^2}{U_e}\frac{x}{\nu}+\frac{2x^2u_*}{U_e\nu}\frac{(-u_*)}{x(\frac{\kappa kU_e}{u_*}+2)}\big)+\\
          = -Re_*^{-2}U_{o3}\big(Re_*+Re_*\frac{-2}{\frac{\kappa U_e}{u_*}+2}\big)+Re_*^{-1}\frac{dU_{o_3}}{dy_o}\big(-y_o+\frac{y_o}{\frac{kU_e}{u_*}+2}\big).
       \end{split}
   \end{equation*}
The $U_{o4}$ term is
\begin{equation}
\frac{d\Delta_3U_{o_4}}{dx}=\frac{d\Delta_3}{dx}U_{o4}+\Delta_3\frac{dU_{o_4}}{dx}=\Delta^{'}_3\frac{Re_x}{dx}U_{o4}+\Delta_3\frac{dU_{o_4}}{dx},
\end{equation}
with the non-dimensional form
   \begin{equation} \label{advection3}
       \begin{split}
       \frac{xU_e}{U_e u_*}U_e u_* \frac{d\Delta_3U_{o4}}{dx}
       =x\Delta^{'}_3\frac{dRe_x}{dx}U_{o_4}+x\Delta_3\frac{dU_{o_4}}{dx}\\
       =x\Delta^{'}_3\frac{U_{o_4}}{U_e\nu}\big(u_x^2+2xu_x\frac{-u_x}{x(\frac{kU_e}{u_x}+2  )}\big)+
       \Delta_3\frac{dU_{o_4}}{dy_o}\big(-y_o+\frac{y_o}{\frac{kU_e}{u_x}+2}\big)\\
       =\Delta^{'}_3Re_xU_{o_4}\big(1+\frac{-2}{\frac{kU_e}{u_x}+2}\big)+\Delta_3\frac{dU_{o_4}}{dy_o}\big(-y_o+\frac{y_o}{\frac{kU_e}{u_x}+2}\big). \\
              \end{split}
       \end{equation}

The non-dimensional advection term due to the velocity defect is
\begin{comment}
\begin{equation} \label{defect_advection}
\begin{split}
 \frac{x}{U_eu_*}(U-U_e)\frac{\partial U}{\partial x}=\frac{x}{U_eu_x}u_*U_o\frac{\partial U}{\partial x}\\
 =\frac{x}{U_e}U_o\Big\{U_{o_1}\frac{du_*}{dx}+u_*\Big[\frac{dU_{o_1}}{dy_o}(-\frac{y_o}{u_*}-\frac{y_o}{x}\frac{du_*}{dx})
  + \frac{1}{U_e}\frac{du_*}{dx}U_{o2}+\Delta_1 \frac{dU_{o2}}{dy_o}(-\frac{y_o}{x}-\frac{y_o}{x}\frac{du_*}{dx})\Big]\Big\}\\
   =-\frac{u_*}{U_e}\frac{U_{o1}^2}{\kappa\frac{U_e}{u_*}+2} + \frac{u_*}{U_e}U_{o1}\Big\{\frac{dU_{o1}}{dy_o}(-y_o+\frac{y_o}{k\frac{u_e}{u_*}+2})
     -\frac{u_*}{U_e}\frac{U_{o2}}{\kappa \frac{U_e}{u_*}+2}    +\frac{u_*}{U_e} \frac{dU_{o2}}{dy_o}(-y_o+\frac{y_o}{k\frac{u_e}{u_*}+2})\Big\}.
 \end{split}
 \end{equation}
 Here we include only terms containing $U_{o1}$ and $U_{o2}$ in the expansion as the latter terms are already of order $u_*^2/U_e^2$. The terms containing
 $Re_*^{-1}$ is of order $(u_*/U_e)Re_*^{-1}$.
\end{comment}
\begin{equation*} \label{eq}
   \frac{x}{U_e u_x}(U-U_e)\frac{\partial U}{\partial x}=\frac{x}{U_e u_*}u_* U_o\frac{\partial U}{\partial x}=\frac{x}{U_e}U_o\frac{\partial U}{\partial x}
   \end{equation*}
   \begin{equation*}
   \begin{split}
   =\frac{x}{U_e}U_o\Big\{(U_{o_1}+\Delta_1 U_{o_2})\frac{du_x}{dx}+u_*\frac{dU_{o_1}+\Delta_1 U_{o_2}}{dx}\\
   +(Re^{-1}_* U_{o_3}+\Delta_2 U_{o_4})\frac{du_*}{dx}+
   u_*\frac{d}{dx}(Re^{-1}_*U_{o_3}+\Delta_3 U_{o_4})\Big\}\\
   =\frac{u_*}{U_e}\big(U_{o_1}+\Delta_1 U_{o_2}+Re^{-1}_* U_{o_3}+\Delta_3 U_{o_4}\big)^2\frac{-1}{k\frac{U_e}{u_x}+2}\\
   +\frac{u_*}{U_e}\Big\{\frac{dU_{o1}}{dy_o}(-y_o+\frac{y_o}{\kappa \frac{U_e}{u_*}+2})-\frac{u_*}{U_e}\frac{U_{o2}}{\kappa \frac{U_e}{u_*}+2}
 +\Delta_1\frac{dU_{o2}}{dy_o}(-y_o+\frac{y_o}{\kappa \frac{U_e}{u_*}+2})\Big\}
   \end{split}
   \end{equation*}
       \begin{equation} \label{defect_advection}
   \begin{split}
   -\frac{u_*}{U_e}U_oRe^{-1}_*U_{o_3}\big(1+\frac{2}{k\frac{U_e}{u_*}+2}\big)+
   \frac{u_*}{U_e}U_oRe^{-1}_*U_{o_3}\frac{dU_{o_3}}{dy_o}\big(-y_o+\frac{y_o}{k\frac{U_e}{u_x}+2} \big).\\
   %+(\frac{u_x}{U_e})^{4}(-2)(U_{o_4}V_o\big(1+\frac{2}{k\frac{U_e}{u_x}+2}\big)\\
   %+(\frac{u_x}{U_e})^{3}U_o\frac{dU_{o_4}}{dy_o}\big(-y_o+\frac{y_o}{k\frac{U_e}{u_x}+2} \big).
    \end{split}
   \end{equation}

 Equation \ref{defect_advection} shows that
   \begin{equation}
\Delta_3=\big(\frac{u_*}{U_e}\big)^{2}=\frac{1}{\ln^2Re_*}. \
   \end{equation}
Thus
   \begin{equation}  
       \Delta_3^{'}=\frac{-2}{\ln^{3} Re_*}\frac{d\ln Re_*}{dRe_*}. \
       =\frac{-2}{\ln^{3} Re_*}\frac{1}{Re_*}.
   \end{equation}
       
The advection due to the normal velocity is obtained as
     \begin{equation}
 \frac{dU}{dy}=u_*\frac{dU_o}{dy}\frac{dy_o}{dy}=u_*\frac{dU_o}{dy_o}\frac{U_e}{u_*x},
 \end{equation}
 %\[V_o=\frac{VU_e}{u_*^2}, \ V=\frac{u_*^2}{U_e}V_o\]
 \begin{equation}
   V\frac{\partial U}{\partial y}=\frac{u_*^2}{U_e}V_o\frac{dU_o}{dy_o}\frac{U_e}{x}=\frac{u_*^2}{x}V_o\frac{dU_o}{dy_o},
 \end{equation}
and
 \begin{equation} \label{V_advection}
 \frac{x}{U_eu_*}\frac{u_*^2}{x}V_o\frac{dU_e}{dy_o}=\frac{u_*}{U_e}V_o\frac{dU_o}{dy_o}.
\end{equation}
The shear-stress derivative is
  \begin{equation}
   \begin{split} \label{shear_derivative}
   \frac{d\overline{uv}}{dy}=u_*^{2}\frac{d\overline{uv}_o}{dy_o}\frac{dy_o}{dy}=u_*^{2}\frac{d\overline{uv_o}}{dy_o}\frac{U_e}{xu_*},\\
   \frac{x}{U_eu_*}\frac{d\overline{uv}}{dy}= \frac{x}{U_eu_*}u_*^{2}\frac{d\overline{uv_o}}{dy_o}\frac{v_e}{xu_*}=\frac{d\overline{uv_o}}{dy_o}.
   \end{split}
   \end{equation}
The viscous term is
\begin{equation} 
   \nu \frac{\partial^{2}U}{\partial y^{2}}=\nu \frac{U_e}{x} \frac{\partial^{2}U_o}{\partial y_o^{2}} \frac{U_e}{xu_*},
   \end{equation}
   \begin{equation} \label{viscous_derivative}
       {\frac{x}{U_e u_*}}\nu{\frac{U_e}{x}}\frac{\partial^{2} U_o}{\partial y_o^{2}}\frac{U_e}{xu_*}
       =\frac{\nu}{xu_*}\frac{U_e}{u_*}\frac{\partial^{2}U_o}{\partial y_o^{2}}
       =\frac{\nu}{\partial u_*}\frac{\partial^{2}U_o}{\partial y_o^2}=Re_*^{-1}\frac{\partial^{2}U_o}{\partial y_o^2}.
   \end{equation}

   The derivative of the normal stress difference is
   \begin{equation}
     \begin{split}
       \frac{\partial u_*^{2}(\overline{u_o^{2}}-\overline{v_o^{2}})}{\partial x}=2u_*\frac{\partial u_*}{dx}(\overline{u_o^{2}}-\overline{v_o^{2})}
       +u_*^{2}\frac{\partial}{\partial x}\big(\overline{u_{o1}^{2}}-\overline{v_{o1}^{2}}+\Delta_1(\overline{u_{o2}^{2}}-\overline{v_{o2}^{2}})+... \big)\\
       =2u_*\frac{\partial u_*}{dx}(\overline{u_o^{2}}-\overline{v_o^{2})}+u_*^{2}\frac{d(\overline{u_{o1}^{2}}-\overline{v_{o1}^{2}})}{d y_o}yU_e\big(\frac{-1}{x^{2}u_*}-{\frac{1}{xu_*^{2}}\frac{\partial u_*}{\partial x}}\big)\\
       +u_*^2\frac{d\Delta_1}{dx}(\overline{u_{o2}^{2}}-\overline{v_{o2}^{2}})
       +u_*^{2}\frac{d(\overline{u_{o2}^{2}}-\overline{v_{o2}^{2}})}{d y_o}yU_e\big(\frac{-1}{x^{2}u_*}-{\frac{1}{xu_*^{2}}\frac{\partial u_*}{\partial x}}\big).
     \end{split}
   \end{equation}
Its non-dimensional form is
   \begin{equation} \label{normal_derivative}
   \begin{split}
     \frac{x}{U_e u_*}\frac{\partial(\overline{u^{2}}-\overline{v^{2}})}{\partial x}=2\frac{x}{U_e}\frac{du_*}{dx}(\overline{u_o^{2}}-\overline{v_o^{2}})
     +\frac{{x}}{U_e{u_*}}u_*^{2}yU_e\big(\frac{-1}{x^{2}u_*}-{\frac{1}{xu_*^{2}}\frac{ du_*}{dx}}\big)\frac{d(\overline{u_{o1}^{2}}-\overline{v_{o1}^{2}})}{dy_o} \\
     +\frac{x}{U_eu_*}u_*^2\frac{-u_*}{U_ex}\frac{\overline{u_{o2}^{2}}-\overline{v_{o2}^{2}}}{\kappa \frac{U_e}{u_*}+2}+
     +\frac{u_*}{U_e}\big(\frac{u_*}{U_e}y_o+\frac{u_*}{U_e}y_o\frac{1}{\kappa\frac{U_e}{u_*}+2}\big)\frac{d(\overline{u_{o2}^{2}}-\overline{v_{o2}^{2}})}{dy_o}\\
       % =-2\frac{x}{U_e}\frac{u_*}{x}\frac{(\overline{u_o^{2}}-\overline{v_o^{2}})}{k\frac{u_o}{u_*}+2}-\big(\frac{u_*y_o}{U_e}+\frac{u_*}{U_e}y_o\frac{x}{u_*}\frac{du_*}{dx}\big)\frac{d(\overline{u_o^{2}}-\overline{v_o^{2}})}{dy_o} \\
        %=-2\frac{u_*}{U_e}\frac{(\overline{u_o^{2}}-\overline{v_o^{2}})}{k\frac{U_e}{u_*}+2}+\big(-\frac{u_*}{U_e}y_o+\frac{u_*}{U_e}y_o\frac{x}{u_*}\frac{U_e}{u_*}\frac{1}{k\frac{U_e}{u_*}+2}\big)\frac{d(\overline{u_o^{2}}-\overline{v_o^{2}})}{dy_o}\\
     =-2\frac{u_*}{U_e}\frac{(\overline{u_o^{2}}-\overline{v_o^{2}})}{k\frac{U_e}{u_*}+2}
     +\frac{u_*}{U_e}\big(-y_o+\frac{1}{k\frac{U_e}{u_*}+2}\big)\frac{d(\overline{u_{o1}^{2}}-\overline{v_{o1}^{2}})}{dy_o}+\frac{U_e^2}{u_*^2}\frac{(\overline{u_{o2}^{2}}-\overline{v_{o2}^{2}})}{k\frac{U_e}{u_*}+2}\\
        +\frac{u_*^2}{U_e^2}\big(-y_o+\frac{1}{k\frac{U_e}{u_*}+2}\big)\frac{d(\overline{u_{o2}^{2}}-\overline{v_{o2}^{2}})}{dy_o}.
        \end{split}
   \end{equation}
   
  Using  \ref{advection1}, \ref{advection2}, \ref{advection3}, \ref{defect_advection}, \ref{V_advection}, \ref{shear_derivative}, \ref{viscous_derivative},
\ref{normal_derivative}, and \ref{full_momentum}, we obtain the  non-dimensional outer momentum equation \ref{U_o_equation}.

\begin{comment}
  \begin{equation}
   \begin{split}
   (-1+\frac{1}{\kappa \frac{U_e}{u_*}+2})y_o\frac{dU_{o1}}{dy_o}-\frac{u_*}{U_e}\frac{U_{o2}}{\kappa \frac{U_e}{u_*}+2}
   +\frac{u_*}{U_e}y_o\frac{dU_{o2}}{dy_o}(-1+\frac{1}{\kappa \frac{U_e}{u_*}+2}) \\
   -Re_*^{-1}U_{o3}\big(1+\frac{-2}{\frac{\kappa U_e}{u_*}+2}\big)+Re_*^{-1}\frac{dU_{o_3}}{dy_o}\big(-y_o+\frac{y_o}{\frac{kU_e}{u_*}+2}\big)\\
   -2\frac{u_*^3}{U_e^3}U_{o_4}\big(1+\frac{-2}{\frac{kU_e}{u_x}+2}\big)+\frac{u_*^2}{U_e^2}\frac{dU_{o_4}}{dy_o}\big(-y_o+\frac{y_o}{\frac{kU_e}{u_x}+2}\big) \\
   -\frac{u_*}{U_e}\frac{U_{o1}^2}{\kappa\frac{U_e}{u_*}+2} + \frac{u_*}{U_e}U_{o1}\Big\{\frac{dU_{o1}}{dy_o}(-y_o+\frac{y_o}{k\frac{u_e}{u_*}+2})
     -\frac{u_*}{U_e}\frac{U_{o2}}{\kappa \frac{U_e}{u_*}+2}    +\frac{u_*}{U_e} \frac{dU_{o2}}{dy_o}(-y_o+\frac{y_o}{k\frac{u_e}{u_*}+2})\Big\}\\
     +\frac{u_*}{U_e}V_o\frac{dU_o}{dy_o}\\
   =-\frac{d\overline{uv_o}}{dy_o}+2\frac{u_*}{U_e}\frac{(\overline{u_o^{2}}-\overline{v_o^{2}})}{k\frac{U_e}{u_*}+2}
     -\frac{u_*}{U_e}\big(-y_o+\frac{1}{k\frac{U_e}{u_*}+2}\big)\frac{d(\overline{u_{o1}^{2}}-\overline{v_{o1}^{2}})}{dy_o}-\frac{U_e^2}{u_*^2}\frac{(\overline{u_{o2}^{2}}-\overline{v_{o2}^{2}})}{k\frac{U_e}{u_*}+2}\\
        -\frac{u_*^2}{U_e^2}\big(-y_o+\frac{1}{k\frac{U_e}{u_*}+2}\big)\frac{d(\overline{u_{o2}^{2}}-\overline{v_{o2}^{2}})}{dy_o}+Re_*^{-1}\frac{\partial^{2}U_o}{\partial y_o^2}
   \end{split}
   \end{equation}
\end{comment}
  
The outer shear-stress budget are obtained as follows.
\begin{equation}
 \begin{split}
\frac{\partial \overline{uv}}{\partial x}=\overline{uv_o}\frac{\partial u_*^{2}}{\partial x}+u_*^{2}\frac{\partial \overline{uv_o}}{\partial x}
=2\overline{uv}_ou_*\frac{-u_*}{x({k\frac{U_e}{u_*}+2})}\\
+u_*^{2}\Big\{\frac{d\overline {uv}_o}{dy_o}yU_e(\frac{-1}{x^2u_*}+\frac{-1}{xu_*^{2}}\frac{du_*}{dx})
+\frac{d\Delta_1}{dx}\overline{uv_{o_2}}+\Delta_1{\frac{d\overline{uv_{o_2}}}{dy_o}yU_e(\frac{-1}{x^2u_*}+\frac{-1}{xu_*^{2}}\frac{du_*}{dx})}\\
+\frac{d}{dx}(Re_*^{-1}\overline{uv}_{o2})+\frac{d\Delta_3\overline{uv}_{o4}}{dx}\Big\}.\\
\end{split}
\end{equation}
Its non-dimensional form up to the order of $\Delta_1$ is
\begin{equation}
  \frac{x}{U_eu_*^{2}}U_e\frac{\partial \overline{uv}}{\partial x}=-\frac{2\overline {uv}_o}{k\frac{U_e}{u_*}+2}
  +\frac{d\overline{uv}_{o1}}{dy_o}y_o(-1+\frac{1}{{k\frac{U_e}{u_*}+2}})- \frac{u_*}{U_e}\frac{\overline{uv}_{o2}}{k\frac{U_e}{u_*}+1}
  +\frac{u_*}{U_e}\frac{d\overline{uv}_{02}}{dy_o}(y_o+\frac{y_o}{k\frac{U_e}{u_*}+2}).
\end{equation}
The order $Re_*$ term is
\begin{equation}
   \begin{split}
       \frac{d}{dx}(Re_*^{-1}\overline{uv}_{o2})=\frac{dRe_*^{-1}}{dx}\overline{uv}_{o3}+Re_*^{-1}\frac{\overline{uv}_{o2}}{dx}\\ 
       =-Re_*^{-2}\frac{dRe_*}{dx}\overline{uv}_{o3}+Re_*^{-1}\frac{d\overline{uv}_{o3}}{dy_o}yU_e(\frac{-1}{x^2u_*}+\frac{-1}{xu_*^2}\frac{du_*}{dx})\\ 
       =-Re_*^{-2}\frac{\overline{uv}_{o3}}{\nu}\frac{1}{U_e}(u^2_*+2xu_*\frac{du_*}{dx})
       +Re_*^{-1}\frac{d\overline{uv}_{o3}}{dy_o}yU_e(\frac{-1}{x^2u_*} +\frac{-1}{xu_*^2}\frac{du_*}{dx}).
       \end{split}
   \end{equation}
   The non-dimensional form is
   \begin{equation}
       \begin{split}
           \frac{x}{U_o u_*}U_eu_*\frac{dRe_x^{-1}\overline{uv}_{o3}}{dx}=-\frac{u_*xRe_*^{-2}}{u_*}\frac{\overline{uv}_{o3}}{\nu}(u_*^2+2xu_*\frac{du_*}{dx})+\\
           \frac{U_exu_*}{U_e u_*}Re_x^{-1}\frac{d\overline{uv}_{o3}}{dy_o}yU_e(\frac{-1}{x^2u_*}+\frac{-1}{xu_*^2}\frac{du_*}{dx})\\
         %  =-Re_x^{-2}U_{o3}\big(\frac{u_*^2}{U_e}\frac{x}{\nu}+\frac{2x^2u_*}{U_e\nu}\frac{(-u_*)}{x(\frac{\kappa kU_e}{u_*}+2)}\big)+\\
          = -Re_x^{-1}\overline{uv}_{o3}\big(1+\frac{-2}{\frac{\kappa U_e}{u_*}+2}\big)+Re_*^{-1}\frac{d\overline{uv}_{o3}}{dy_o}\big(-y_o+\frac{y_o}{\frac{kU_e}{u_*}+2}\big).
       \end{split}
   \end{equation}
The order $\Delta_3$ term is
   \begin{equation}
     \frac{d\Delta_3\overline{uv}_{o4}}{dx}=\frac{d\Delta_3}{dx}\overline{uv}_{o4}+\Delta_3\frac{d\overline{uv}_{o4}}{dx}=\Delta^{'}_3\frac{dRe_x}{dx}\overline{uv}_{o4}
       +\Delta_3\frac{d\overline{uv}_{o4}}{dx},
\end{equation}
with the non-dimensional form
   \begin{equation}
       \begin{split}
       \frac{xU_e}{U_e u_*}U_e u_* \frac{d\Delta_3\overline{uv}_{o4}}{dx}
       =x\Delta^{'}_3\frac{dRe_x}{dx}\overline{uv}_{o4}+x\Delta_3\frac{d\overline{uv}_{o4}}{dx}\\
       =x\Delta^{'}_3\frac{\overline{uv}_{o4}}{U_e\nu}\big(u_x^2+2xu_x\frac{-u_x}{x(\frac{kU_e}{u_x}+2  )}\big)+
       \Delta_3\frac{d\overline{uv}_{o4}}{dy_o}\big(-y_o+\frac{y_o}{\frac{kU_e}{u_x}+2}\big)\\
       =-2\frac{u_*^3}{U_e^3}\overline{uv}_{o4}\big(1+\frac{-2}{\frac{kU_e}{u_x}+2}\big)+\frac{u_*^2}{U_e^2}\frac{d\overline{uv}_{o4}}{dy_o}\big(-y_o+\frac{y_o}{\frac{kU_e}{u_x}+2}\big). \\
              \end{split}
       \end{equation}

   The advection due to the velocity defect is

\begin{equation}
     \begin{split}
       \frac{x}{U_eu_*^{2}}(U-U_e)\frac{\partial \overline{uv}}{\partial x}=-\frac{u_*}{U_e}\frac{2U_o\overline{uv_o}}{k\frac{U_o}{u_*}+2}
       +\frac{u_*}{U_e}U_o\frac{d\overline{uv}_{o2}}{dy_o}y_o(-1+\frac{1}{\frac{kU_e}{u_x}+2})+(\frac{u_*}{u_e})^{2}U_{o_1}\frac{-2\overline{uv}_{o2}}{\frac{kU_e}{u_x}+2}\\
       -\frac{u_*^2}{U_e^2}U_{o1}\frac{d\overline{uv}_{o2}}{dy_o}(-1+\frac{-1}{\frac{kU_e}{u_x}+2}).\\
          \end{split}
 \end{equation}
The advection due to the normal velocity is
\begin{equation}
     \begin{split}
     \frac{\partial \overline{uv}}{\partial y}=u_*^{2}\frac{\partial \overline{uv_o}}{\partial y_o}\frac{U_e}{xu_*}, \ \
     V\frac{\partial \overline{uv}}{\partial y}=\frac{u_*^{2}}{U_e}V_o u_*^{2}\frac{\partial \overline{uv_o}}{\partial y_o}\frac{U_e}{xu_*}\\
     \frac{x}{U_eu_*^{2}}V\frac{\partial \overline{uv}}{\partial y}=\frac{u_*}{U_e}V_o\frac{\partial \overline{uv_o}}{\partial y_o}.\\
     \end{split}
 \end{equation}
The shear-stress production is
\begin{equation}
     \begin{split}
     \overline {v_o^{2}}\frac{\partial U}{\partial y}=u_*^{2}v_o^{2}u_*\frac{\partial U_o}{\partial y_o}\frac{U_e}{xu_*},\\
     \frac{x}{U_eu_*^{2}}\overline{v^{2}}\frac{\partial U}{\partial y}=\overline{v_o^{2}}\frac{\partial U_o}{\partial y_o}.
     \end{split}
 \end{equation}

The viscous diffusion is
 \begin{equation}
     \begin{split}
       \frac{\partial^{2}\overline{uv}}{\partial y^{2}}=u_*^{2}\frac{\partial^{2}\overline{uv_o}}{dy_o^{2}}(\frac{U_e}{xu_*^{2}}),\\
       \frac{x}{U_eu_*^{2}}\nu\frac{\partial^{2}\overline{uv}}{\partial y^{2}}=\frac{x\nu}{U_e u_*^{2}}\frac{U_e^{2}}{x^{2}}\frac{d^{2}\overline{uv}_o}{dy_o^{2}}
       =\frac{\nu}{u_*x}\frac{U_e}{u_*}\frac{\partial^{2}\overline{uv}_o}{dy_o^{2}}
       =\frac{\nu}{U_e\delta}\frac{U_e}{u_*}\frac{\partial^{2}\overline{uv_o}}{dy_o^{2}}=\frac{\nu}{u_*\delta}\frac{\partial^{2}\overline{uv}_o}{dy_o^{2}}.
     \end{split}
 \end{equation}

 Combining the above terms we obtain the outer shear-stress budget \ref{shear_budget}.
\begin{comment}
 \begin{equation}
  \begin{split}
    -\frac{2\overline {uv}_o}{k\frac{U_o}{u_*}+2}
  +\frac{d\overline{uv}_{o1}}{dy_o}y_o(-1+\frac{1}{{k\frac{U_e}{u_*}+2}})- \frac{u_*}{U_e}\frac{\overline{uv}_{o2}}{k\frac{U_o}{u_*}+1}
  +\frac{u_*}{U_e}\frac{d\overline{uv}_{02}}{dy_o}(y_o+\frac{y_o}{k\frac{U_o}{u_*}+2})\\
  -Re_*^{-1}\overline{uv}_{o3}\big(1+\frac{-2}{\frac{\kappa U_e}{u_*}+2}\big)+Re_*^{-1}\frac{d\overline{uv}_{o3}}{dy_o}\big(-y_o+\frac{y_o}{\frac{kU_e}{u_*}+2}\big)\\
  -2\frac{u_*^3}{U_e^3}\overline{uv}_{o4}\big(1+\frac{-2}{\frac{kU_e}{u_x}+2}\big)+\frac{u_*^2}{U_e^2}\frac{d\overline{uv}_{o4}}{dy_o}\big(-y_o+\frac{y_o}{\frac{kU_e}{u_x}+2}\big) \\
  -\frac{u_*}{U_e}\frac{2U_o\overline{uv_o}}{k\frac{U_o}{u_*}+2}
       +\frac{u_*}{U_e}U_o\frac{d\overline{uv}_{o2}}{dy_o}y_o(-1+\frac{1}{\frac{kU_e}{u_x}+2})+(\frac{u_*}{u_e})^{2}U_{o_1}\frac{-2\overline{uv}_{o2}}{\frac{kU_e}{u_x}+2}\\
       -\frac{u_*^2}{U_e^2}U_{o1}\frac{d\overline{uv}_{o2}}{dy_o}(-1+\frac{-1}{\frac{kU_e}{u_x}+2})+\frac{u_*}{U_e}V_o\frac{\partial \overline{uv_o}}{\partial y_o}\\
       =-\overline{\Big(u\frac{\partial p}{\partial y}+v\frac{\partial p}{\partial x}\Big)}_o -\overline{v_o^{2}}\frac{\partial U_o}{\partial y_o}+Re_*^{-1}\frac{\partial^{2}\overline{uv}_o}{dy_o^{2}}\\
     \end{split}
 \end{equation}
\end{comment}

The perturbation momentum equation for the inner is derived as follows.
  \begin{equation}
 \begin{split}
   \frac{\partial U}{\partial x}=\frac{\partial u_* U_i}{\partial x}=\frac{d u_*}{d x}U_i
   +u_*\frac{\partial(U_{i1}+{\delta_1}U_{i2}+....)}{\partial x} \\
   =\frac{d u_*}{d x}U_i+ u_*\Big\{\frac{dU_{i1}}{dy^+}\frac{dy^+}{dx}+\frac{d\delta_1}{dx}U_{i2}+ \delta_1\frac{dU_{i2}}{dy^+}\frac{dy^+}{dx}+...\Big\}\\
 =\frac{d u_*}{d x}U_i+\frac{u_*y}{\nu}\big(\frac{du_*}{dx}\big(\frac{dU_{i1}}{dy^+}+\delta_1\frac{dU_{i2}}{dy^+}+...\big)+u_*\frac{d\delta_1}{dx}U_{i2},\\
 \end{split}
 \end{equation}
 where $dy^+/dx=(y/\nu)(du_*/dx)$. The non-dimensional advection due to $U$ is
\begin{equation}
     \begin{split}
       \frac{\delta_\nu}{u_*^{2}}U\frac{\partial U}{\partial x}=\frac{\nu}{u_*^{3}}u_*U_i
       \Big\{\frac{d u_*}{d x}U_i+\frac{u_*y}{\nu}\big(\frac{du_*}{dx}\big(\frac{dU_{i1}}{dy^+}+\delta_1\frac{dU_{i2}}{dy^+}+...\big)+u_*\frac{d\delta_1}{dx}U_{i2}\Big\}\\
         =\frac{\nu}{u_*^{{2}}}U_i\Big\{\frac{-{u_*}U_i}{x(k\frac{U_e}{u_*}+2)}+y^{+}\frac{{-u_*}}{x(k\frac{U_e}{u_*}+2)}\big(\frac{dU_{i1}}{dy^+}+\delta_1\frac{dU_{i2}}{dy^+}+...\big)\Big\}\\
       %  =\frac{\nu U_i}{u_*}\big(-\frac{U_i}{\frac{\delta}{\delta_0}\frac{u_{*_0}}{u_*}x_0}-y^+\frac{1}{\frac{\delta}{\delta_0}\frac{u_{*_0}}{u_*}x_0}\big(\frac{dU_{\Omega_1}}{dy_\Omega}+\epsilon_\Omega\frac{dU_{\Omega_2}}{dy_2}+...\big)\frac{1}{k\frac{U_e}{u_*}+2}\\
\sim \Big\{-\frac{\nu}{U_e \delta}U_i^{2}-\frac{\nu}{U_e\delta}y^+U_i\big(\frac{dU_{i1}}{dy+}+\delta_1\frac{dU_{i2}}{dy^+}+...\big)\Big\}\frac{1}{k\frac{U_e}{u_*}+2}\\
\sim Re_\delta^{-1}\frac{1}{k\frac{U_e}{u_*}+2}  \Big\{-U_i^{2}-y^+U_i\big(\frac{dU_{i1}}{dy+}+\delta_1\frac{dU_{i2}}{dy^+}+...\big)\Big\},         
     \end{split}
 \end{equation}
where $\nu/(u_* x) \approx \nu/(U_e\delta )$ has been used. The shear-stress derivative is
\begin{equation}
  \frac{\partial \overline{uv}}{\partial y}=u_*^{2}\frac{\partial \overline{uv}_i}{\partial y^+}\frac{u_*}{\nu}.
  \end{equation}
The advection due to $V$ is
\begin{equation}
\frac{\partial U}{\partial y}=u_*\frac{\partial U_i}{\partial y^+}\frac{u_*}{\nu}, \ \ \frac{\partial^{2}U}{\partial y^2}=u_*\frac{\partial^{2}U_i}{\partial y^{+2}}(\frac{u_*}{\nu})^{2},
\end{equation}
\begin{equation}
  V\frac{\partial U}{\partial y}=\frac{\nu}{\delta} \frac{u_*}{U_e}u_*V_i\frac{u_*^{2}}{\nu}\frac{\partial U_i}{\partial y^{+}}, \ \
  \frac{\delta_\nu}{u_*^{2}}V\frac{\partial U}{\partial y}=\frac{\nu}{\delta U_e}V_i\frac{\partial U_i}{\partial y^{+}}. 
\end{equation}
The normal-stress difference derivative is
\begin{equation}
     \begin{split}
       \frac{\partial \overline{u^{2}}-\overline{v^{2}}}{\partial x}=\frac{\partial u_* U_e \overline{u_i^{2}}}{\partial x}-\frac{\partial u_*^{2}\overline{v_i^{2}}}{\partial x}=\frac{du_*}{dx}U_e \overline{u_i^{2}}+u_*U_e\frac{\partial\overline{u_i^{2}}}{\partial x}
       -\big(-2u_*\frac{du_*}{dx}\overline{v_i^{2}}+u_*^{2}\frac{\partial \overline{v_i^{2}}}{\partial x}\big)\\
       =-\frac{u_*}{x}\frac{1}{k\frac{U_e}{u_*}+2}U_e\overline{u_i^{2}}+u_*U_e\frac{\partial \overline{u_i^{2}}}{\partial y^+}\frac{y}{\nu}\frac{d u_*}{dx}
       -\big(2u_*\frac{u_*}{x}\frac{\overline{v_i^{2}}}{k\frac{U_e}{u_*}+2}+u_*^{2}\frac{d \overline{v_i^{2}}}{dy^{+}}\frac{y}{\nu}\frac{du_*}{dx}\big)\\
       =-\frac{u_*}{x}\frac{1}{k\frac{U_e}{u_*}+2}U_e\overline{u_i^{2}}
       -u_*U_e\frac{y}{\nu}\frac{u_*}{x}\frac{1}{k\frac{U_e}{u_*}+2}\frac{\partial \overline{u_i^{2}}}{\partial y^{+}}
       -\big(-2u_*\frac{u_*}{x}\frac{\overline{v_i^{2}}}{k\frac{U_e}{u_*}+2}-u_*^{2}\frac{\partial \overline{u_i^{2}}}{\partial y^{+}}\frac{y}{\nu}\frac{u_*}{x}\frac{1}{k\frac{U_e}{u_*}+2}\big),
     \end{split}
\end{equation}
where the scaling $\overline{u^{2}}\sim u_*^{2}\ln Re_*\sim u_*U_e$ (\citealt{Townsend76}) has been used. Its non-dimensional form is 
 \begin{equation}
     \begin{split}
       \frac{\delta_\nu}{u_*^{2}}\frac{\partial \overline{u^{2}}-\overline{v^{2}}}{\partial x}=
       -\frac{\nu}{u_*^{2}}\frac{U_e}{x}\frac{\overline u_i^{2}}{k\frac{U_e}{u_*}+2}
       -\frac{\nu U_e}{u_*^{2}}\frac{y}{\nu}\frac{u_*}{x}\frac{1}{k\frac{U_e}{u_*}+2}\frac{\partial \overline{u_i^{2}}}{\partial y^{+}}
       -\Big\{-2\frac{\nu}{u_*}{x}\frac{v_i^{2}}{k\frac{U_e}{u_*}+2}-\frac{\nu}{u_*}\frac{y}{\nu}\frac{u_*}{x}\frac{1} {k\frac{U_e}{u_*}+2}\frac{\partial \overline{v_i^{2}}}{\partial y^{+}}\Big\}\\
       =-\frac{\nu}{u_*\delta}\frac{ \overline{u_i^{2}}}{k\frac{U_e}{u_*}+2}
       -\frac{\nu}{u_*\delta}\frac{1}{k\frac{U_e}{u_*}+2}y^{+}\frac{d \overline{u_i^{2}}}{d y^{+}}
       -\Big\{-2\frac{\nu}{U_e \delta}\frac{\overline{v_i^{2}}}{k\frac{U_e}{u_*}+2}-\frac{\nu}{U_e\delta}\frac{y^{+}}{k\frac{U_e}{u_*}+2}\frac{\partial \overline{v_i^{2}}}{\partial y^{+}}\Big\}\\
       \approx -Re_\delta^{-1}\overline{u_i^{2}}-Re_\delta^{-1}y^{+}\frac{d \overline{u_i^{2}}}{d y^{+}}+
       2Re_\delta^{-1}\frac{u_*}{U_e}\overline{v_i^{2}}+Re_\delta^{-1}\frac{u_*}{U_e}y^{+}\frac{\partial \overline{v_i^{2}}}{\partial y^{+}}.
     \end{split}
 \end{equation}

 The inner perturbation equation up to the order of $Re_\delta^{-1}u_*/U_e$ is
\begin{equation}
     \begin{split}             
       -Re_\delta^{-1}\frac{u_*}{U_e}\Big\{U_i^{2}+y^{+}\frac{dU_i}{dy^{+}}\Big\}+Re_\delta^{-1}V_i \frac{dU_i}{dy^{+}}
       =-\frac{\partial \overline {uv_i}}{\partial y^{+}}+\frac{d^{2}U_i}{dy^{+2}}\\
       -Re_\delta^{-1}\overline{u_i^{2}}-Re_\delta^{-1}y^{+}\frac{d \overline{u_i^{2}}}{d y^{+}}+
       2Re_\delta^{-1}\frac{u_*}{U_e}\overline{v_i^{2}}+Re_\delta^{-1}\frac{u_*}{U_e}y^{+}\frac{\partial \overline{v_i^{2}}}{\partial y^{+}}.
%       -\frac{\nu}{u_*\delta}\frac{ \overline{u_i^{2}}}{k\frac{U_e}{u_*}+2} \\
 %      -\frac{\nu}{u_*\delta}\frac{1}{k\frac{U_e}{u_*}+2}y^{+}\frac{d \overline{u_i^{2}}}{d y^{+}}
  %     +\Big\{2\frac{\nu}{U_e \delta}\frac{\overline{v_i^{2}}}{k\frac{U_e}{u_*}+2}+\frac{\nu}{U_e\delta}\frac{y^{+}}{k\frac{U_e}{u_*}+2}\frac{\partial \overline{v_i^{2}}}{\partial y^{+}}\Big\}
     \end{split}
 \end{equation}

 The inner shear-stress budget is obtained as follows.
 \begin{equation}
     \begin{split}
       \frac{\partial \overline{uv}}{\partial x}=2u_*\frac{du_*}{dx}\overline{uv}_i+u_*^{2}\frac{d \overline{uv}_i}{d y^{+}}\frac{dy^{+}}{dx}
       =2u_*\frac{u_*}{x}\frac{-1} {k\frac{U_e}{u_*}+2}\overline{uv_i}+u_*^2\frac{d \overline{uv_i}}{d y^{+}}\frac{y}{\nu}\frac{du_*}{dx},
%u_*^{2}\frac{\partial \overline{uv_i}}{\partial y^{+}}y^{+}\frac{1} {k\frac{u_*}{U_e}+2}\\
     \end{split}
 \end{equation}
 \begin{equation}
     \begin{split}
       \frac{\delta_\nu}{u_*^{3}}U\frac{\partial \overline{uv}}{\partial x}
       =\frac{\delta_\nu}{u_*^{3}}u_*U_i\frac{2u_*^{2}}{x}\frac{-\overline{uv_i}}{(k\frac{u_*}{U_e}+2)}
       +\frac{\delta_\nu}{u_*^{3}}u_*^3U_i\frac{d\overline {uv_i}}{dy^{+}}y^{+}\frac{1}{x}\frac{-1}{(k\frac{u_*}{U_e}+2)}\\ 
       =\frac{\delta_\nu u_*}{\delta U_e}U_i\frac{\overline{uv_i}}{k\frac{U_e}{u_*}+2}
       +\frac{\delta_\nu u_*}{\delta U_e}U_iy^{+}\frac{\frac{d\overline{uv_i}}{dy^{+}}}{k\frac{U_e}{u_*}+2}\\
    % \delta U_e\sim xu_*\\
     =\frac{\nu}{U_e\delta}U_i\big(\frac{\overline{uv_i}}{k\frac{U_e}{u_*}+2}+y^{+}\frac{d \overline{uv_i}}{d y^{+}}\big).
     \end{split}
 \end{equation}
 The advection term due to $V$ is
 \begin{equation}
     \begin{split}
         V\frac{\partial \overline{uv}}{\partial y}=\frac{\delta_\nu u_*}{\delta U_e}u_*V_iu_*^{2}\frac{\partial \overline{uv_i}}{\partial y^{+}}\frac{u_*}{\nu},\\
         \frac{\delta_\nu}{u_*^{3}}V\frac{\partial \overline{uv}}{\partial y}
         =\frac{\delta_\nu}{u_*^{3}}\frac{\delta_\nu}{\delta}\frac{u_*}{U_e}u_*^{3}\frac{u_*}{\nu}V_i\frac{d \overline{uv_i}}{d y^{+}},\\
         =\frac{\nu}{U_e\delta}V_i\frac{d \overline{uv_i}}{d y^{+}}.
 \end{split}
 \end{equation}
The production is
 \begin{equation}
     \begin{split}
         \overline {v^{2}}\frac{\partial U}{\partial y}=u_*^{2}\overline{v_i^{2}}u_*\frac{\partial U_i}{\partial y^{+}}\frac{u_*}{\nu}, \ \
         \frac{\delta \nu}{u_*^{3}}\overline {v_i^{2}}\frac{\partial U_i}{\partial y^{+}}\frac{1}{\delta \nu}=\overline {v_i^{2}}\frac{\partial U_i}{\partial y^{+}}.
         %\frac{\partial^{2}\overline{uv}}{\partial x^{2}}=2\big(2u_*\frac{\partial u_*}{\partial x}\frac{1}{x}\frac{1} {k\frac{U_e}{u_*}+2}-2\frac{u_*^{2}}{x^{2}}\frac{1}{k..+2}\frac{u_*^{2}}{x}\frac{\frac{-kU_e}{u_*^{2}}}{{(k...)^{2}}}\frac{du_*}{dx}\big)\overline{uv_i}+\\
     %4u_*\frac{du_*}{dx}\frac{d\overline{uv_i}}{dy^{+}}\frac{y}{\nu}\frac{d^{2}u_*}{dx^{2}}\\
     \end{split}
 \end{equation}
 The viscous diffusion is
 \begin{equation}
     \begin{split}
         \frac{\partial \overline{ uv}}{\partial y}=u_*^{2}\frac{\partial \overline {uv_i}}{\partial y^{+}}\frac{u_*}{\nu},\ \
         \frac{\partial^{2}\overline{uv}}{\partial y^{2}}=u_*^{2}\frac{\partial^{2}\overline{uv_i}}{dy^{+2}}\big(\frac{u_*}{\nu}\big)^{2},\\
         \frac{\delta v}{u_*^{3}}\nu\frac{\partial^{2}\overline{uv}}{dy^{2}}=\frac{\delta\nu}{u_*{3}}\nu\frac{u_*^{4}}{\nu^{2}}\frac{\partial^{2}\overline{uv_i}}{\partial y^{+2}}=\frac{\partial^{2}\overline{uv_i}}{\partial y^{+2}}.
     \end{split}
 \end{equation}

The inner shear-stress budget up to the order of $Re_\delta^{-1}u_*/ U_e$ is obtained as

\begin{equation}
     \begin{split}      
       -Re_\delta^{-1}U_i\big(\frac{\overline{uv_i}}{k\frac{U_e}{u_*}+2}+y^{+}\frac{d \overline{uv_i}}{d y^{+}}\big)+Re_\delta^{-1}V_i\frac{d \overline{uv_i}}{d y^{+}}
       =-\overline{\Big(u\frac{\partial p}{\partial y}+v\frac{\partial p}{\partial x}\Big)}_i \\
       -\overline {v_i^{2}}\Big(\frac{d U_{i1}}{d y^{+}}+Re_\delta^{-1}\frac{d U_{i2}}{d y^{+}}+Re_\delta^{-1}\frac{u_*}{\kappa U_e}\frac{d U_{i3}}{d y^{+}}\Big)
       +\frac{d^{2}\overline{uv}_{i1}}{d y^{+2}}+Re_\delta^{-1}\frac{d^{2}\overline{uv}_{i2}}{d y^{+2}}
       +Re_\delta^{-1}\frac{u_*}{\kappa U_e}\frac{d^{2}\overline{uv}_{i3}}{d y^{+2}}.
     \end{split}
 \end{equation}

\bibliographystyle{jfm}
\bibliography{clemson1}
\end{document}